\documentclass[lettersize,journal]{IEEEtran}
\usepackage{amsmath,amsfonts}
\usepackage{algorithm}
\usepackage{algpseudocode}
\usepackage{array}
\usepackage[caption=false]{subfig}
\usepackage{textcomp}
\usepackage{stfloats}
\usepackage{url}
\usepackage{verbatim}
\usepackage{graphicx}
\usepackage{cite}
\usepackage{multirow}

\usepackage{soul}
\usepackage{xcolor}
\usepackage{todonotes}
\usepackage[colorlinks=true,linkcolor=blue,urlcolor=blue,urlcolor=blue,citecolor=blue]{hyperref}

\usepackage{amssymb}

\begin{document}

\definecolor{lightgray}{gray}{0.9}
\definecolor{lightblue}{rgb}{0.9,0.9,1}
\definecolor{LightMagenta}{rgb}{1,0.5,1}
\definecolor{red}{rgb}{1,0,0}
\definecolor{brightgreen}{rgb}{0.4, 1.0, 0.0}

\newcommand\couldremove[1]{{\color{lightgray} #1}}
\newcommand{\remove}[1]{}
\newcommand{\move}[2]{ {\textcolor{Purple}{ \bf --- MOVE #1: --- }} {\textcolor{Orchid}{#2}} }

\newcommand{\hlc}[2][yellow]{ {\sethlcolor{#1} \hl{#2}} }
\newcommand\note[1]{\hlc[SkyBlue]{-- #1 --}} 

\newcommand\mynote[1]{\hlc[yellow]{#1}}
\newcommand\tingjun[1]{\hlc[yellow]{TC: #1}}
\newcommand\zhihui[1]{\hlc[LightMagenta]{ZG: #1}}
\newcommand\tom[1]{\hlc[brightgreen]{TOM: #1}}
\newcommand\andy[1]{\hlc[pink]{Andy: #1}}
\newcommand\bill[1]{\hlc[cyan!50]{Bill: #1}}
\newcommand\change[1]{{\color{blue} {#1}}}

\newcommand{\TODO}[1]{\textcolor{red}{#1}}
\newcommand{\revise}[1]{\textcolor{blue}{#1}}


\newcommand{\myparatight}[1]{\vspace{0.5ex}\noindent\textbf{#1~}}

\newcommand{\cmark}{\ding{51}}%
\newcommand{\xmark}{\ding{55}}%
\newcommand{\greencheck}{\color[HTML]{3C8031}{\cmark}}
\newcommand{\redcross}{\color[HTML]{ED1B23}{\xmark}}
\newcommand{\greenno}{\color[HTML]{3C8031}{\textbf{No}}}
\newcommand{\redyes}{\color[HTML]{ED1B23}{\textbf{Yes}}}
\newcommand{\greenlow}{\color[HTML]{3C8031}{\textbf{Low}}}
\newcommand{\redhigh}{\color[HTML]{ED1B23}{\textbf{High}}}

\newcommand*\circledwhite[1]{\tikz[baseline=(char.base)]{
            \node[shape=circle,draw,inner sep=0.6pt] (char) {\scriptsize{#1}};}}

\newcommand*\circled[1]{\tikz[baseline=(char.base)]{
            \node[shape=circle,draw,fill=black,text=white,inner sep=1pt] (char) {\scriptsize{#1}};}}

\newcommand{\name}{{SS-CGA}}
\newcommand{\namebf}{{\textbf{SS-CGA}}}

\newcommand{\agora}{Agora}
\newcommand{\agorabf}{\textbf{Agora}}

\newcommand{\armavec}{\sf Savannah-mc (arma-vec)}
\newcommand{\armacube}{\sf Savannah-mc (arma-cube)}

\newcommand{\specialcell}[2][c]{%
  \begin{tabular}[#1]{@{}c@{}}#2\end{tabular}}

\newcommand{\iu}{{j}}

\newcommand{\littlesum}{\mathop{\textstyle\sum}}
\newcommand{\littleint}{\mathop{\textstyle\int}}

\newcommand{\siso}{SISO}
\newcommand{\mimoTwoByTwo}{2$\times$2}
\newcommand{\mimoFourByFour}{4$\times$4}

\newcommand{\bbdev}{\textsf{bbdev}}

\newcommand{\scs}{{\Delta f}}
\newcommand{\scNum}{N_{\textrm{sc}}}
\newcommand{\sampRate}{F_{\textrm{s}}}
\newcommand{\fftSize}{N_{\textrm{FFT}}}
\newcommand{\chMat}{{\textbf{H}}}
\newcommand{\chVec}{{\textbf{h}}}
\newcommand{\precodeMat}{{\textbf{P}}}

\newcommand{\usec}{$\mu$s} 
\newcommand{\msec}{ms}     

\newcommand{\fft}{\textsf{fft}}
\newcommand{\ifft}{\textsf{ifft}}
\newcommand{\csi}{\textsf{csi}}
\newcommand{\precode}{\textsf{precode}}
\newcommand{\encode}{\textsf{enc}}
\newcommand{\decode}{\textsf{dec}}
\newcommand{\modul}{\textsf{modul}}
\newcommand{\demod}{\textsf{demod}}
\newcommand{\equal}{\textsf{equal}}

\newcommand{\tbSize}{T}
\newcommand{\tbCrcSize}{T_{\textrm{crc}}}
\newcommand{\cbSize}{K_{\textrm{cb}}}
\newcommand{\cbNum}{N_{\textrm{cb}}}
\newcommand{\liftingSize}{Z_{c}}
\newcommand{\liftingSizeSet}{\mathbf{\Theta}}
\newcommand{\fillerBitNum}{N_{\textrm{filler}}}

\newcommand{\codeRate}{R}

\newcommand{\throughput}{Tp}
\newcommand{\codingTime}{t}
\newcommand{\informationBits}{K'}

\newcommand{\thresPower}{\theta_{\textrm{Power}}}
\newcommand{\thresOtsu}{\theta_{\textrm{Otsu}}}
\newcommand{\thresPSD}{\theta_{\textrm{PSD}}}
\newcommand{\thresIoU}{\theta_{\textrm{IoU}}}

\newcommand{\tput}{\eta} 
\newcommand{\lat}{l} 

\newcommand{\se}{E}

\newcommand{\numDelayBin}{M}
\newcommand{\numDopplerBin}{N}
\newcommand{\delay}{\tau}
\newcommand{\DopplerShift}{\nu}
\newcommand{\delayRes}{\Delta\delay} 
\newcommand{\DopplerRes}{\Delta\DopplerShift} 
\newcommand{\freqSpacing}{\Delta f} 
\newcommand{\pathIdx}{p}
\newcommand{\pathDelay}{\tau_{\pathIdx}}
\newcommand{\pathDoppler}{\nu_{\pathIdx}}
\newcommand{\delayIdx}{k}
\newcommand{\DopplerIdx}{l}
\newcommand{\pilotDelayIdx}{K_0}
\newcommand{\pilotDopplerIdx}{L_0}
\newcommand{\pathDelayIdx}{{\delayIdx}_{\pathIdx}}
\newcommand{\pathDopplerIdx}{{\DopplerIdx}_{\pathIdx}}

\newcommand{\chEleDD}{H_{\textrm{dd}}}
\newcommand{\chMatDD}{\textbf{H}_{\textrm{dd}}}
\newcommand{\chEleDDEst}{\widehat{H}_{\textrm{dd}}}
\newcommand{\chMatDDEst}{\widehat{\textbf{H}}_{\textrm{dd}}}
\newcommand{\chMatDDEstHerm}{\widehat{\textbf{H}}^\textrm{H}_{\textrm{dd}}}

\newcommand{\chMatDDRowIdx}{q}
\newcommand{\chMatDDColIdx}{r}
\newcommand{\chMatDDColIdxMapped}[1]{{\chMatDDColIdx}_{\pathIdx}(#1)}
\newcommand{\chMatDDColIdxMappedRowIdx}{\chMatDDColIdxMapped{\chMatDDRowIdx}}
\newcommand{\chMatDDRowIdxMapped}[1]{{\chMatDDRowIdx}_{\pathIdx}(#1)}
\newcommand{\chMatDDRowIdxMappedColIdx}{\chMatDDRowIdxMapped{\chMatDDColIdx}}
\newcommand{\sparseCoe}{D}
\newcommand{\mvmCoe}[2]{\sparseCoe_{{#1},{#2}}}
\newcommand{\coePathRowIdx}{\mvmCoe{\pathIdx}{\chMatDDRowIdx}}
\newcommand{\chMatDDColDelayBinOffset}{{d}_{\delayIdx}(\pathIdx)}
\newcommand{\chMatDDColDopplerBinOffset}{{d}_{\DopplerIdx}(\pathIdx)}
\newcommand{\chMatDDRowDelayBinOffset}{{\delta}_{\delayIdx}(\pathIdx)}
\newcommand{\chMatDDRowDopplerBinOffset}{{\delta}_{\DopplerIdx}(\pathIdx)}
\newcommand{\chPathGain}{h_{\pathIdx}}
\newcommand{\phCompFactor}{\gamma_{\pathIdx,{\chMatDDRowIdx}}}
\newcommand{\phTwisted}{\phi_{\pathIdx,{\chMatDDRowIdx}}}
\newcommand{\rawDelayOffset}{A_{\pathIdx,{\chMatDDRowIdx}}}
\newcommand{\DelayOffsetWrapCnt}{n_{\pathIdx,{\chMatDDRowIdx}}}

\newcommand{\chNoiseDDVec}{\textbf{w}_{\textrm{dd}}}
\newcommand{\chNoiseDDMat}{\textbf{W}_{\textrm{dd}}}
\newcommand{\chNoiseDDMatEle}{W_{\textrm{dd}}}
\newcommand{\chNoiseTD}{w}

\newcommand{\chEleMMSE}{H_{\textrm{LMMSE}}}
\newcommand{\chMatMMSE}{\textbf{H}_{\textrm{LMMSE}}}

\newcommand{\chResponse}{h_{\textrm{eff}}}

\newcommand{\chEffEstMat}{\widehat{\textbf{h}}_{\textrm{eff}}}
\newcommand{\chEffEstMatEle}{\widehat{h}_{\textrm{eff}}}
\newcommand{\chEffMat}{\mathbf{h}_{\textrm{eff}}}
\newcommand{\chEffMatEle}{h_{\textrm{eff}}}

\newcommand{\sigTxDD}{X_{\textrm{dd}}}
\newcommand{\sigTxDDVec}{\mathbf{x}_{\textrm{dd}}}
\newcommand{\sigTxDDVecEle}{x_{\textrm{dd}}}
\newcommand{\sigTxDDVecEst}{\widehat{\mathbf{x}}_{\textrm{dd}}}
\newcommand{\sigTxDDVecEstDemodEle}{\breve{x}_{\textrm{dd}}}
\newcommand{\sigTxDDMatEle}{X_{\textrm{dd}}}
\newcommand{\sigTxDDMat}{\mathbf{X}_{\textrm{dd}}}
\newcommand{\sigTxDDMatEst}{\widehat{\mathbf{X}}_{\textrm{dd}}}

\newcommand{\sigTxTdVecEle}{x}
\newcommand{\sigTxTdVec}{\mathbf{x}}
\newcommand{\sigTxTdMatEle}{X}
\newcommand{\sigTxTdMat}{\mathbf{X}}

\newcommand{\sigRxDD}{Y_{\textrm{dd}}}
\newcommand{\sigRxDDVec}{\mathbf{y}_{\textrm{dd}}}
\newcommand{\sigRxDDMatEle}{Y_{\textrm{dd}}}
\newcommand{\sigRxDDMat}{\mathbf{Y}_{\textrm{dd}}}
\newcommand{\sigRxTd}{y}
\newcommand{\sigRxTdVec}{\mathbf{y}}
\newcommand{\sigRxTdMat}{\mathbf{Y}}

\newcommand{\chMaxDopplerFreq}{\nu_{\text{max}}}

\newcommand{\bandwidth}{B}
\newcommand{\frameTime}{T}
\newcommand{\numPath}{P} 
\newcommand{\coherenceTime}{T_c}

\newcommand{\heffThres}{\theta}
\newcommand{\chPathThresSet}{\widehat{\Omega}}

\newcommand{\noiseCovMat}{\mathbf{R}_{n}}
\newcommand{\cgaIter}{\xi}
\newcommand{\cgaIters}{\Xi}
\newcommand{\sqResNorm}{c_\textrm{norm}}

\newcommand{\kernelFftMatEle}{E_{\text{Zak}}}
\newcommand{\kernelFftMat}{\mathbf{E}_{\text{Zak}}}
\newcommand{\twistMatEle}{E_{\text{twist}}}
\newcommand{\twistMat}{\mathbf{E}_{\text{twist}}}

\newcommand{\delayIdxSet}{\mathcal{\numDelayBin}}
\newcommand{\DopplerIdxSet}{\mathcal{\numDopplerBin}}
\newcommand{\pathIdxSet}{\mathcal{\numPath}}
\newcommand{\gridFlatIdxSet}{\mathcal{Q}} 

\newcommand{\jimg}{j}
\newcommand{\defn}{\triangleq}
\newcommand{\mymod}[2]{\langle #1 \rangle_{#2}}

\newcommand{\myAbs}[1]{\left|{#1}\right|}
\newcommand{\myAng}[1]{\angle{#1}}
\newcommand{\myConjugate}[1]{{#1}^{*}}
\newcommand{\myTranspose}[1]{{#1}^{\top}}
\newcommand{\myHermitian}[1]{{#1}^{H}}
\newcommand{\myIsFunc}[1]{\mathbf{1}\{#1\}}

\newcommand{\AoD}{\phi}
\newcommand{\AoDVec}{\bm{\upphi}}
\newcommand{\AoDbf}{\boldsymbol\phi}
\newcommand{\AoDDirectional}{\Phi}
\newcommand{\az}{\phi}
\newcommand{\azVec}{\bm{\upphi}}
\newcommand{\azVecUE}{\bm{\upphi}_{\textrm{UE}}}
\newcommand{\azbf}{\boldsymbol\phi}
\newcommand{\el}{\psi}
\newcommand{\elbf}{\boldsymbol\psi}

\newcommand{\ElemComp}{w}
\newcommand{\ElemCompbf}{\mathbf{w}}
\newcommand{\ElemCompNew}{w^\prime}
\newcommand{\ElemCompNewbf}{\mathbf{w}^\prime}
\newcommand{\ElemAmp}{A}
\newcommand{\ElemAmpbf}{\mathbf{A}}
\newcommand{\ElemPhase}{\theta}
\newcommand{\ElemPhasebf}{\boldsymbol\theta}
\newcommand{\steer}{s}
\newcommand{\steerVec}{\mathbf{s}}
\newcommand{\steermat}{\mathbf{S}}
\newcommand{\beamPattern}{BP}

\newcommand{\bw}{B}
\newcommand{\carrierFreq}{f_{c}}
\newcommand{\carrierWave}{\lambda}

\newcommand{\csiMat}{\mathbf{H}}

\newcommand{\ASA}[2]{\textrm{ASA}({#1},{#2})}
\newcommand{\antNum}{N}
\newcommand{\antIdx}{n}
\newcommand{\antDist}{d}
\newcommand{\subarrayNum}{M}
\newcommand{\subarraySet}{\mathcal{M}}
\newcommand{\subarrayIdx}{m}
\newcommand{\subarrayAntNum}{N_{s}}
\newcommand{\subarrayAntIdx}{n}
\newcommand{\subarrayAntDist}{d}

\newcommand{\setSubarray}{\mathcal{A}}
\newcommand{\subarrayAlloc}{a}
\newcommand{\subarrayAllocVec}{\mathbf{a}}
\newcommand{\subarrayAllocMat}{\mathbf{A}}
\newcommand{\subarrayAllocSet}{\mathbb{A}}

\newcommand{\bfWeight}{w}
\newcommand{\bfWeightVec}{\mathbf{w}}
\newcommand{\bfAmp}{A}
\newcommand{\bfAmpVec}{\mathbf{A}}
\newcommand{\bfPhase}{\theta}
\newcommand{\bfPhaseVec}{\boldsymbol{\theta}}
\newcommand{\bfGain}{g}
\newcommand{\bfGainSig}[1]{g^{\textrm{sig}}_{#1}}
\newcommand{\bfGainInt}[2]{g^{\textrm{int}}_{{#1}\rightarrow{#2}}}

\newcommand{\power}{\mathcal{P}}
\newcommand{\powerSignal}{\mathcal{S}}
\newcommand{\powerSignalDiff}{d\mathcal{S}}
\newcommand{\powerInterf}{\mathcal{I}}
\newcommand{\powerInterfDiff}{d\mathcal{I}}
\newcommand{\powerNoise}{N}

\newcommand{\userNum}{U}
\newcommand{\userIdx}{u}
\newcommand{\userSet}{\mathcal{U}}

\newcommand{\userNumSub}{K}

\newcommand{\userSelected}{k}
\newcommand{\userSelectedNum}{K}
\newcommand{\userSelectedSet}{\mathcal{K}}

\newcommand{\userAngle}{\phi}
\newcommand{\userWeight}{\alpha}

\newcommand{\baseSNR}{\gamma}
\newcommand{\SNR}{\mathsf{SNR}}
\newcommand{\SNRMax}{\mathsf{SNR}^{\textrm{max}}}
\newcommand{\SINR}{\mathsf{SINR}}
\newcommand{\SINRMax}{\mathsf{SINR}^{\textrm{max}}}
\newcommand{\Capacity}{T}
\newcommand{\Rate}{R}
\newcommand{\RateMax}{\Rate^{\textrm{max}}}
\newcommand{\RateAvg}{\widebar{\Rate}}
\newcommand{\CapacityMax}{\Tilde{T}}
\newcommand{\suppress}{\alpha}

\newcommand{\RateHist}{\widebar{\Rate}}

\newcommand{\past}{p}
\newcommand{\decay}{\beta}

\newcommand{\RateMean}{\Bar{R}}
\newcommand{\JFI}{\mathsf{JFI}}

\newcolumntype{+}{>{\global\let\currentrowstyle\relax}}
\newcolumntype{^}{>{\currentrowstyle}}
\newcommand{\rowstyle}[1]{%
  \gdef\currentrowstyle{#1}%
  #1\ignorespaces
}

\newenvironment{spmatrix}[1]
 {\def\mysubscript{#1}\mathop\bgroup\begin{bmatrix}}
 {\end{bmatrix}\egroup_{\textstyle\mathstrut\mysubscript}}

\newcommand{\bigO}{\mathcal{O}} 
\newcommand{\conv}{\ast}

\title{Real-Time and Scalable Zak-OTFS Receiver Processing on GPUs}

\author{Junyao Zheng$^{*}$, Chung-Hsuan Tung$^{*}$, Yuncheng Yao, Nishant Mehrotra, Sandesh Mattu, Zhenzhou Qi, \\ Danyang Zhuo, Robert Calderbank, and Tingjun Chen
\thanks{This work was supported in part by NSF under grants CNS-2211944, CCF-2342690, ECCS-2434131, CNS-2443137, CNS-2450567, and OAC-2503010, the AFOSR under grants FA 8750-20-2-0504 and FA 9550-23-1-0249, and an NVIDIA Academic Grant.
$^{*}$indicates equal contribution.}
\thanks{
J. Zheng, C.-H. Tung, N. Mehrotra, S. Mattu, Z. Qi, R. Calderbank, and T. Chen are with the Department of Electrical and Computer Engineering, Duke University, NC 27708, USA (email: \{junyao.zheng, chunghsuan.tung, tingjun.chen\}@duke.edu).
Y. Yao and D. Zhuo are with the Department of Computer Science, Duke University, NC 27708, USA.}
\vspace{-10mm}
}


\maketitle

\begin{abstract}
Orthogonal time frequency space (OTFS) modulation offers superior robustness to high-mobility channels compared to conventional orthogonal frequency-division multiplexing (OFDM) waveforms.
However, its explicit delay-Doppler (DD) domain representation incurs substantial signal processing complexity, especially with increased DD domain grid sizes.
To address this challenge, we present a scalable, real-time Zak-OTFS receiver architecture on GPUs through hardware–algorithm co-design that exploits DD-domain channel sparsity.
Our design leverages compact matrix operations for key processing stages, a branchless iterative equalizer, and a structured sparse channel matrix of the DD domain channel matrix to significantly reduce computational and memory overhead.
These optimizations enable low-latency processing that consistently meets the 99.9-th percentile real-time processing deadline.
The proposed system achieves up to {906.52}\thinspace{Mbps} throughput with a DD grid size of (16384,32) using 16QAM modulation over {245.76}\thinspace{MHz} bandwidth.
Extensive evaluations under a Vehicular-A channel model demonstrate strong scalability and robust performance across CPU (Intel Xeon) and multiple GPU platforms (NVIDIA Jetson Orin, RTX 6000 Ada, A100, and H200), highlighting the effectiveness of compute-aware Zak-OTFS receiver design for next-generation (NextG) high-mobility communication systems.
\end{abstract}


\section{Introduction}
\IEEEPARstart{O}{rthogonal} time frequency space (OTFS) modulation has received significant attention in NextG networks due to its robustness on channel Doppler shift compared to traditional orthogonal frequency-division multiplexing (OFDM) modulation.
OTFS was first proposed as multi-carrier OTFS (MC-OTFS)~\cite{hadani2017mcotfs} by adding an inverse symplectic finite Fourier transform (ISFFT) precoder to existing OFDM architectures for compatibility, but it suffers from spectral efficiency loss due to the cyclic prefix (CP).
In contrast, Zak-OTFS~\cite{mohammed2022otfs,mohammed2023otfs} directly uses the Zak transform to map the delay-Doppler (DD) domain symbols to time-domain pulse-train waveforms, i.e., \emph{pulsones}, that span the entire frame duration, enhancing channel predictability and enabling more structured processing.

Since OTFS modulates information symbols as a 2D grid in the DD domain, its baseband digital signal processing (DSP) involves matrix operations that incur high computational complexity, especially with increased DD grid sizes.
Hence, recent research efforts have focused on developing low-complexity algorithms and practical OTFS systems, as summarized in Table~\ref{tab:existing_otfs_proc_sys}. 
Several studies have investigated hardware implementations of OTFS modulation and demodulation using software-defined radios (SDRs)~\cite{thaj2019otfs,wei2025otfs,lin2026enabling,nauman2025otfs}, graphics processing units (GPUs)~\cite{yap2023}, or field-programmable gate arrays (FPGAs)~\cite{jou2025,wang2025,isik2023fpga,shadangi2024vlsi,dora2023low}.
These implementations typically focus on demonstrating the functional feasibility of OTFS systems.

Despite extensive work on OTFS algorithms and system prototypes, the \emph{scalability} and \emph{real-time processing} perspectives of OTFS systems remain largely underexplored.
Most prior evaluations consider small-to-moderate DD grid sizes~\cite{thaj2019otfs, zheng2025zak, mattu2025low, wei2025otfs, nauman2025otfs}, ranging from tens to a few hundred bins in the delay ($\numDelayBin$) and Doppler ($\numDopplerBin$) domains, primarily due to the high complexity of OTFS detection algorithms.
As a result, OTFS signal processing latency grows rapidly and becomes prohibitive with large grid sizes (e.g., $\numDelayBin$ or $\numDopplerBin$ on the order of several to more than ten thousand).
In addition, while several proof-of-concept prototypes have demonstrated functional OTFS systems~\cite{thaj2019otfs, wei2025otfs, lin2026enabling, nauman2025otfs, zheng2025zak}, performing the \emph{full OTFS receiver processing in real time at frame rate}--i.e., the DSP of a frame should be completed within the frame duration with a high satisfaction rate--remains challenging.
Achieving such real-time processing is challenging for large DD grids, where operations such as channel estimation and equalization involve high-dimensional arithmetic computations.

\begin{table*}[!t]
    \centering
    \caption{Recent OTFS processing systems with full receiver pipeline.}
    \vspace{-1.0mm}
    \renewcommand{\arraystretch}{1.15}
    \begin{tabular}{|c|c|c|c|c|c|c|c|}
    \hline
     & \cite{thaj2019otfs} (2019) & \cite{yap2023} (2023) & \cite{jou2025} (2025) & \cite{wang2025} (2025) & \cite{zheng2025zak} (2025) & \cite{mattu2025low} (2026) & \textbf{This Work}\\
    \hline
    \textbf{Compute Platform} 
    & CPU & GPU (A100) & FPGA (Alveo U250) & CPU & CPU & CPU & \textbf{GPU (H200)} \\
    \hline
    \textbf{OTFS Formulation} & MC-OTFS & MC-OTFS & MC-OTFS & MC-OTFS & Zak-OTFS & Zak-OTFS & \textbf{Zak-OTFS} \\
    \hline
    \textbf{DD Grid Size}, $(\numDelayBin, \numDopplerBin)$ 
    & (32, 32) & (156, 60) & (2048, 8) & (2048, 512) & (31, 37) & (31, 37) & \textbf{(16384, 32)} \\
    \hline
    \textbf{Bandwidth}, $\bandwidth$ (MHz) 
    & 25 & 5 & 40 & 300 & 0.93 & 0.93 & \textbf{245.76} \\
    \hline
    \textbf{Frame Duration}, $\frameTime$ ({\msec})  
    & 1.28 & 1.000 & 0.461 & 3.495 & 1.233 & 1.233 & \textbf{1.067} \\
    \hline
    \textbf{Equalizer} 
    & MP & MP & MRC & MRC & LMMSE & CGA  & \textbf{SS-CGA} \\
    \hline
    \textbf{Data Rate} 
    & 3.20 Mbps & 6.00 Mbps & 280.32 Mbps & 1.80 Gbps & 1.86 Mbps & 1.86 Mbps & \textbf{906.52 Mbps}\\
    \hline
    \textbf{Real-Time Processing}  
    & No & Yes & Yes & No & No & No & \textbf{Yes} \\
    \hline
    \end{tabular}
    \label{tab:existing_otfs_proc_sys}
    \vspace{-2.0mm}
\end{table*}

These computational characteristics make GPUs a promising platform for real-time OTFS DSP.
Modern GPU architectures are optimized for high-throughput parallel computation and provide specialized hardware and software support for operations such as fast Fourier transforms, matrix-vector multiplications (MVM), and general matrix multiplication (GEMM).
Importantly, GPUs also provide extremely high on-package memory bandwidth, which is critical when manipulating large DD tensors and channel operators during equalization.
As the OTFS grid size increases, the receiver workload increasingly resembles large-scale structured linear algebra, a regime in which GPUs significantly outperform traditional CPUs due to their massive parallelism and memory throughput.
At the same time, GPU programmability enables rapid exploration of evolving OTFS receiver algorithms, which is particularly valuable given the active research landscape.
However, fully exploiting GPU-based OTFS processing requires careful hardware-algorithm co-design.

In this paper, \emph{we design and implement a scalable, GPU-based real-time Zak-OTFS signal processing system}, guided by three key insights arising from the unique coupling between the Zak-OTFS receiver DSP and GPU processing.
First, many OTFS receiver DSP stages involve compact matrix operations with coefficients that depend solely on system parameters such as grid size and index mappings. 
Examples include phase factors associated with the Zak transform and channel estimation in the DD domain.
Precomputing these coefficients eliminates repeated transcendental or indexing-heavy computations, enabling efficient matrix operations during runtime.

Second, the wireless channel typically exhibits a limited number of paths, resulting in strong sparsity in the DD domain effective channel, $\chEffMat \in \mathbb{C}^{\numDelayBin\times\numDopplerBin}$.
While the DD domain channel matrix, $\chMatDD$, can be written as an ${\numDelayBin}{\numDopplerBin} \times {\numDelayBin}{\numDopplerBin}$ matrix, it is directly derived from $\chEffMat$ with inherent structured sparsity.
Efficient OTFS receiver implementations should therefore avoid explicit construction of dense matrices and instead exploit representations that are aware of such structured sparsity (SS).
By leveraging MVM with sparse matrices for equalization, the computational complexity and memory footprint can be significantly reduced while aligning with GPU computation.

Third, the computational efficiency of a GPU is significantly undermined by branch conditions, e.g., if-else checks.
However, many low-complexity equalizers rely on iteration and conditional termination.
Hence, a branchless iteration strategy, rather than dynamic early termination, is necessary.
To address this, our implementation determines the optimal number of iterations through comprehensive profiling.
By replacing runtime branching with this profile-based approach, the receiver ensures a seamless processing pipeline with peak utilization while maintaining the desired bit-error rate (BER).

We implement the real-time Zak-OTFS processing system in Python with PyTorch and customized Triton kernels for GPU, along with a C++ baseline for CPU.
Through comprehensive evaluations across four NVIDIA GPU (Jetson Orin, RTX~6000 Ada, A100, and H200) and an Intel CPU (Xeon 6348) platforms, our design meets the 99.9-th percentile processing latency deadline of {2.13}\thinspace{\msec} for the DD domain grid size of up to $(16384, 32)$, yielding a throughput of {906.52}\thinspace{Mbps} with 16QAM modulation and {245.76}\thinspace{MHz} bandwidth.
In summary, this paper makes the following key contributions:
\begin{itemize}
    \item We leverage compact matrix operation, exploit the sparsity of the DD domain channel, and propose a branchless low-complexity equalizer for efficient Zak-OTFS receiver processing on GPUs;
    \item We implement a practical, scalable computing system for real-time OTFS signal processing that supports an OTFS DD grid size up to $(16384, 32)$;
    \item We conduct a comprehensive evaluation in terms of end-to-end processing latency and BER across various computing platforms and wireless channel conditions.
\end{itemize}

\emph{To the best of our knowledge, this is the first implementation that
(i) investigates and combines compute hardware- and channel sparsity-awareness for OTFS baseband processing; and
(ii) achieves real-time OTFS processing for DD domain grid sizes of up to $(\numDelayBin, \numDopplerBin)=(16384,32)$.}
The code, benchmarks, and profiling results are open-sourced at~\cite{otfs-github}.

\section{Related Work}
\label{sec:related}

\myparatight{Complexity of OTFS processing.}
Since OTFS operates on a 2D resource grid in the DD domain and each impulse in the grid is cross-dependent, channel estimation and equalization are of high complexity.
Point pilots~\cite{mohammed2023otfs} and spread pilots~\cite{ubadah2024zak} are two pilot designs in Zak-OTFS, where channel estimation based on the point pilot involves simple element-wise computation in $\bigO(\numDelayBin\numDopplerBin)$, and that on the spread pilot relies on the cross-ambiguity function in $\bigO(\numDelayBin^2\numDopplerBin^2)$.
Reducing the complexity of equalization has received significant attention~\cite{Surabhi2020, chong2025cross, mishra2021otfs}.
These methods perform equalization in near-linear complexity ($\bigO(\numDelayBin\numDopplerBin\log{\numDelayBin\numDopplerBin})$) by exploiting the block circulant structure of the modulation.
Iterative equalizers such as message-passing (MP)~\cite{raviteja2018interference}, maximum ratio combining (MRC)~\cite{thaj2020mrc}, and conjugate gradient algorithm (CGA)~\cite{mattu2025low} can further reduce complexity to linear per-iteration.

\myparatight{Practical OTFS system implementations.}
Table~\ref{tab:existing_otfs_proc_sys} compares this work against the related literature across different compute platforms, link parameters, the equalizer algorithm, system throughput, and the real-time capability.
The CPU has been used for algorithm prototyping and complexity-reduction studies~\cite{thaj2019otfs, zheng2025zak, mattu2025low, Zhang2025_lowcompl_zakotfs, thaj2020mrc, mattu2025differential}, e.g., the CGA equalizer~\cite{mattu2025low} is evaluated on a $(31, 37)$ DD domain grid on the CPU.
Similarly, the MP equalizer has been implemented on a $(32, 32)$ grid~\cite{thaj2019otfs}, while the MRC equalizer has been evaluated on a larger $(2048, 512)$ grid~\cite{wang2025}.
However, these CPU-based implementations do not report their real-time processing capability.
As the theory of OTFS matures, efforts to implement OTFS processing across a variety of computing platforms aim to enable practical real-time operation under different bandwidth and grid configurations~\cite{yap2023, jou2025, wang2025, zheng2025zak}.
Many implementations leverage the FPGA platforms~\cite{jou2025, isik2023fpga, dora2023low, shadangi2024vlsi} for their ability to provide deterministic latency and energy-efficient hardware acceleration.
For example, a real-time OTFS receiver on an Alveo U250 FPGA with an MRC equalizer achieves a throughput of {280.32\thinspace{Mbps} with a $(2048, 8)$ grid~\cite{jou2025}.
In contrast, GPU-based real-time OTFS systems remain relatively limited despite their potential for large-scale parallel processing, with only a proof-of-concept demonstration on an NVIDIA A100 using an MP equalizer for a $(156, 60)$ grid~\cite{yap2023}.
Hence, further investigation leveraging GPU's massive parallelism for real-time processing of high-bandwidth OTFS signal is in demand.

\myparatight{Virtualized RAN (vRAN) systems leveraging GPUs.}
Recent literature reveals interest in using GPUs to support vRAN systems for compute-intensive baseband processing of large matrices.
For example, while Agora~\cite{ding2020agora} provides a MIMO baseband processing framework for 5G using CPU and Savannah~\cite{qi2024savannah} accelerates it to stringent frequency range 2 (FR2) deadlines, MegaStation~\cite{xie2025building} demonstrates a reduced tail latency and improved throughput on massive MIMO by migrating the system onto a supercomputer with pooled GPUs.
In addition, DecodeX~\cite{qi2026decodex} benchmarks LDPC decoding workloads, which are considered the most compute-intensive stage of the receiver pipeline, across CPU/GPU/ASIC, and demonstrates the potential benefit of a GPU-resident pipeline for efficient resource sharing in vRAN.
Similarly, X5G~\cite{villa2025x5g} integrates NVIDIA Aerial with OpenAirInterface in an end-to-end private 5G O-RAN testbed, where GPUs accelerate PHY-layer functions and achieve multi-Gbps downlink throughput.
The success of GPU-based OFDM processing in 5G systems reveals its potential to accelerate the processing pipeline for advanced waveforms such as OTFS in NextG networks.

\section{Preliminaries}
\label{sec:prelim}

In this paper, we consider the Zak-OTFS system, shown in Fig.~\ref{fig:otfs_pipeline}, which adopts a twisted convolution representation in the DD domain.
Key receiver DSP stages include the Zak transform, effective channel computation ($\chEffMat$), DD channel construction ($\chMatDD$), equalization, and demodulation, many of which introduce large structured linear algebra workloads.

\begin{figure*}[!t]
    \centering
    \includegraphics[width=0.95\linewidth]{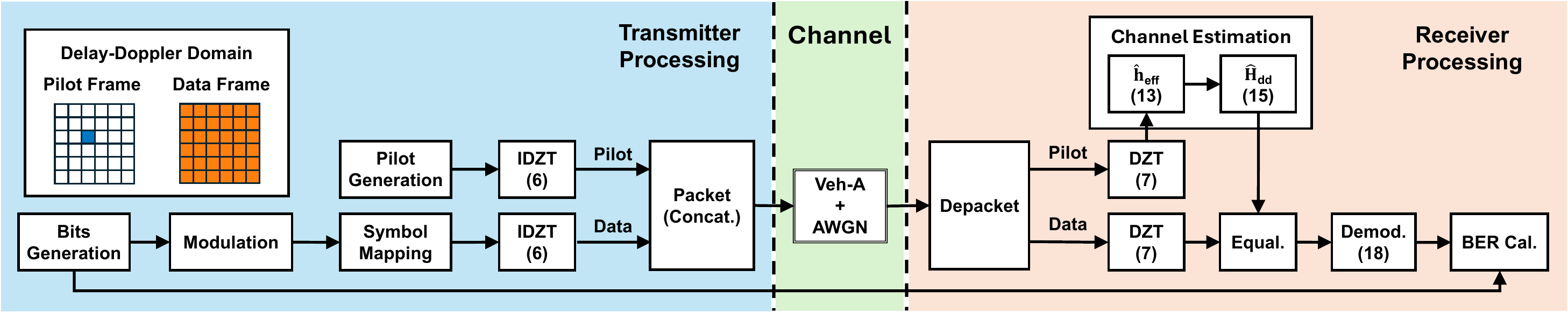}
    \vspace{-2.0mm}
    \caption{Zak-OTFS signal processing pipeline based on discrete Zak transform (DZT). A point-pilot frame and a data frame are concatenated to form a packet. The simulated channel considers Veh-A~\cite{itur_m1225} for paths with delay and Doppler shifts, and AWGN for SNR adjustment. The receiver pipeline stages include DZT, channel estimation, equalization, and demodulation, which will be discussed in Section~\ref{subsec:prelim_pipeline}.}
    \label{fig:otfs_pipeline}
    \vspace{-2.0mm}
\end{figure*}

\subsection{Zak-OTFS Modulation and Parameters}
\label{subsec:prelim_otfs}

In Zak-OTFS, $\numDelayBin\numDopplerBin$ information symbols, ${\sigTxDDMatEle}[{\delayIdx},{\DopplerIdx}]$, are multiplexed over a frame consisting of an $\numDelayBin \times \numDopplerBin$-dimensional DD grid with $\numDelayBin$ delay bins and $\numDopplerBin$ Doppler bins.
Let
$\delayIdxSet \defn \{0, 1, \dots, M-1\}$ and $\DopplerIdxSet \defn \{0, 1, \dots, N-1\}$, this DD domain information grid can be denoted by
\begin{equation}
    \left(
        \delayIdx \cdot \delayRes,
        \DopplerIdx \cdot \DopplerRes
    \right),\
    \delayIdx \in \delayIdxSet,\
    \DopplerIdx \in \DopplerIdxSet,
\end{equation}
with a delay and Doppler period of $\numDelayBin\cdot\delayRes$ and $\numDopplerBin\cdot\DopplerRes$, respectively, where $\delayRes$ and $\DopplerRes$ are the delay and Doppler resolutions.
$\delayRes$ and $\DopplerRes$ are given by
\begin{equation}\label{eq:delay_Doppler_res}
\delayRes = {1}/{\bandwidth}\
\text{and}\
\DopplerRes = {1}/{\frameTime},
\end{equation}
where $\bandwidth$ is the signal bandwidth (which determines the sampling rate) and $\frameTime$ is the frame duration,
\begin{equation}\label{eq:bandwidth_frametime}
    \bandwidth = \numDelayBin \cdot \freqSpacing\
    \text{and}\
    \frameTime = {\numDopplerBin}/{\freqSpacing},
\end{equation}
with $\freqSpacing$ the frequency spacing.
Hence,
\begin{equation}
    \bandwidth = \numDelayBin (\numDopplerBin\cdot\DopplerRes)\
    \text{and}\
    \frameTime = \numDopplerBin (\numDelayBin\cdot\delayRes).
\end{equation}
The DD domain grid size ($\numDelayBin\times\numDopplerBin$) as well as the delay and Doppler resolutions ($\delayRes$ and $\DopplerRes$) are designed so that
\begin{equation}
    {(\numDelayBin \delayRes)}\cdot{(\numDopplerBin \DopplerRes)} = 1\
    \text{and}\
    {\bandwidth}{\frameTime} = {\numDelayBin}{\numDopplerBin}.
\end{equation}

\subsection{Zak-OTFS Channel and Processing Pipeline}
\label{subsec:prelim_pipeline}

Fig.~\ref{fig:otfs_pipeline} illustrates the Zak-OTFS TX/RX processing pipeline and channel simulation considered in this work.
This section provides the formulation of each critical DSP stage, with a focus on the receiver pipeline and stage complexity.

After data bits are generated and modulated into the delay--Doppler (DD) domain, each symbol of a data packet will be mapped onto a 2D grid and transformed into the time domain using the inverse discrete Zak transform (IDZT)~\eqref{eq:iZak}.
In contrast, the pilot frame is generated directly in the DD domain and converted to the time domain using IDZT.
A packet contains one pilot frame and a number of data frames, depending on the channel coherence time~\cite{mohammed2023otfs}.
Vehicular-A (Veh-A)~\cite{itur_m1225} model is applied to simulate a channel with multiple path delays and Doppler shifts, and we use AWGN on top of Veh-A as the channel noise (details in Section~\ref{sec:impl}).

The receiver pipeline consists of both pilot- and data-frame processing, including the discrete Zak transform (DZT) {\eqref{eq:zak}},
channel estimation including obtaining the channel response grid $\chEffEstMat$ {\eqref{eq:heff}} and constructing the channel matrix $\chMatDDEst$,
followed by equalization (with LMMSE, MRC, and CGA equalizers) and demodulation {\eqref{eq:demod}}.
Finally, the demodulated bits are compared with the ground-truth bits to calculate BER.

\subsubsection{\underline{Zak Transform}}

OTFS waveform constructs the resource grid in the DD domain, and its conversion to (from) the time-domain waveform is via the Zak (inverse Zak) transform.

\myparatight{Inverse discrete Zak transform (IDZT).}
After the pilot and data frames are generated as DD grids, the transmitter converts the DD domain signal matrix $\sigTxDDMat \in \mathbb{C}^{\numDelayBin \times \numDopplerBin}$ to a time-domain signal vector $\sigTxTdVec \in \mathbb{C}^{\numDelayBin \numDopplerBin}$ using IDZT~\cite[Eq.~(13)]{lampel2022otfs},
\begin{equation}
\label{eq:iZak}
{\sigTxTdVecEle}[i]
=
\frac{1}{\sqrt{\numDopplerBin}}
\sum_{\DopplerIdx=0}^{\numDopplerBin-1}
    {\sigTxDD}[\mymod{i}{\numDelayBin},{\DopplerIdx}]
    \cdot
    e^{\jimg 2\pi \frac{\left\lfloor i/\numDelayBin \right\rfloor}{\numDopplerBin}\DopplerIdx},\ \forall i \in {\gridFlatIdxSet},
\end{equation}
where $\mymod{\cdot}{\numDelayBin}$ denotes the modulo $\numDelayBin$ operation and ${\gridFlatIdxSet} \defn \{0,1,\ldots,{\numDelayBin}{\numDopplerBin}-1\}$.
$\sigTxTdVec$ is transmitted through the channel (discussed in Section~\ref{sub2sec:_prelim_channel}), and results in the received time-domain signal vector $\sigRxTdVec \in \mathbb{C}^{{\numDelayBin}{\numDopplerBin}}$.

\myparatight{Discrete Zak transform (DZT).}
The receiver converts $\sigRxTdVec \in \mathbb{C}^{\numDelayBin\numDopplerBin}$ back into the DD domain as matrix $\sigRxDDMat \in \mathbb{C}^{{\numDelayBin}\times{\numDopplerBin}}$ using DZT~\cite[Eq.~(1)]{lampel2022otfs} for all ${\delayIdx} \in {\delayIdxSet}$ and ${\DopplerIdx} \in {\DopplerIdxSet}$:
\begin{equation}
\label{eq:zak}
{\sigRxDD}[{\delayIdx},{\DopplerIdx}]
=
\frac{1}{\sqrt{\numDopplerBin}}
\sum_{i=0}^{{\numDopplerBin}-1}{\sigRxTd}
    [\delayIdx+i{\numDelayBin}]
    \cdot
    e^{-\jimg 2\pi \frac{{i\DopplerIdx}}{{\numDopplerBin}}}.
\end{equation}
Note that both DZT and IDZT reshape the signal vector to/from matrix when transforming the signal domain.

\subsubsection{\underline{Effective Channel in the DD Domain}}\label{sub2sec:_prelim_channel}

Zak-OTFS models the time-varying channel as the DD domain response
\begin{equation}\label{eq:h_tau_nu}
h(\delay,\DopplerShift)
=
\sum_{\pathIdx=0}^{\numPath-1}
    h_{\pathIdx}
    \delta(\delay - \delay_\pathIdx)
    \delta(\DopplerShift - \DopplerShift_\pathIdx),
\end{equation}
where $\numPath$ denotes the number of resolvable paths in the DD domain, $h_{\pathIdx}$ is the complex-valued channel gain of the $\pathIdx$-th path, $\delta(\cdot)$ denotes the Dirac delta function, and $\tau_{\pathIdx}$ and $\nu_\pathIdx$ are the delay and Doppler shift associated with the $\pathIdx$-th path, respectively.
By associating $\tau_{\pathIdx}$ and $\nu_\pathIdx$ with the DD grid
\begin{equation}
\tau_{\pathIdx} = {\delayIdx_{\pathIdx}}/{\bandwidth} = \delayIdx_{\pathIdx}{\delayRes},
\quad
\nu_\pathIdx = {\DopplerIdx_\pathIdx}/{\frameTime} = {\DopplerIdx_\pathIdx}{\DopplerRes},
\end{equation}
the discrete form of the channel model \eqref{eq:h_tau_nu} is denoted as the \emph{effective channel}, $\chEffMat = \left[ \chEffMatEle[{\delayIdx},{\DopplerIdx}] \right] \in \mathbb{C}^{{\numDelayBin}\times{\numDopplerBin}}$, where
\begin{equation}\label{eq:heff_model}
\chEffMatEle[\delayIdx,\DopplerIdx]
=
\sum_{\pathIdx=0}^{\numPath-1}
    h_\pathIdx
    \delta[\delayIdx-\delayIdx_\pathIdx]
    \delta[\DopplerIdx-\DopplerIdx_\pathIdx].
\end{equation}
where $\delta[\cdot]$ denotes the Kronecker delta function, and ${\delayIdx_{\pathIdx}}$ and  ${\DopplerIdx_\pathIdx}$ are the delay and Doppler shift indices on the DD grid for the $\pathIdx$-th path, respectively.

The receiver samples the continuous signal undergoing \eqref{eq:heff_model} at each time point $t = \tfrac{m}{\bandwidth}$, $m \in \delayIdxSet$, and results in the discrete time-domain signal $\sigRxTdVec$ that
\begin{equation}\label{eq:y_discrete_td}
\sigRxTd[i]
\hspace{-0.5mm} = \hspace{-1.5mm}
\sum_{\pathIdx=1}^{\numPath}
    h_{\pathIdx}\hspace{-1mm}
    \sum_{n=0}^{{\numDelayBin}{\numDopplerBin}-1}\hspace{-1mm}
        \sigTxTdVecEle[n]
        \delta[\tfrac{i-n}{\bandwidth}-\tau_{\pathIdx}]
        e^{j2\pi\delay_{\pathIdx}\big(\tfrac{i}{\bandwidth}-\delay_\pathIdx\big)}+
        \chNoiseTD[i],
\end{equation}
where $\delta[\cdot]$ denotes the Kronecker delta function and $\chNoiseTD[\cdot]$ is the additive noise.
Converting \eqref{eq:y_discrete_td} to DD domain, each element $\sigRxDDMatEle[{\delayIdx},{\DopplerIdx}]$ of the received OTFS frame $\sigRxDDMat$ is expressed as a discrete twisted convolution $*_{\sigma_{{}_{d}}}$\footnote{
    $
    a[{\delayIdx},{\DopplerIdx}]
    *_{\sigma_{{}_{d}}}
    b[{\delayIdx},{\DopplerIdx}]
    \defn
    \sum_{{\delayIdx}',{\DopplerIdx}'\in\mathbb{Z}}
        a[{\delayIdx}-{\delayIdx}',{\DopplerIdx}-{\DopplerIdx}']
        b[{\delayIdx}',{\DopplerIdx}']
        e^{\frac{\jimg2\pi}{\numDelayBin\numDopplerBin}{\delayIdx}'({\DopplerIdx}-{\DopplerIdx}')}
    $
}
between the quasi-periodic extension of $\sigTxDDMatEle[{\delayIdx},{\DopplerIdx}]$\footnote{
    $
    \sigTxDDMatEle[{\delayIdx}+n\numDelayBin,{\DopplerIdx}+m\numDopplerBin]
    =
    \sigTxDDMatEle[{\delayIdx},{\DopplerIdx}]
    e^{\jimg\tfrac{2\pi}{\numDopplerBin}n{\DopplerIdx}}
    $, with $
    {\delayIdx} \in \delayIdxSet$,
    ${\DopplerIdx} \in \DopplerIdxSet$,
    $m,n\in\mathbb{Z}
    $
} and the effective channel $\chEffMatEle[{\delayIdx},{\DopplerIdx}]$, i.e.,
\begin{equation}\label{eq:y_channel_DD}
\sigRxDD[\delayIdx,\DopplerIdx]
=
\chEffMatEle[\delayIdx,\DopplerIdx]
*_{\sigma_{{}_{d}}}
\sigTxDD[\delayIdx,\DopplerIdx]
+
\chNoiseDDMatEle[\delayIdx,\DopplerIdx],\
\forall \delayIdx, \DopplerIdx,
\end{equation}
where $\chNoiseDDMat$ is the DD domain additive noise.

\subsubsection{\underline{Channel Estimation}}
To capture the input/output relationship of the channel, a frame containing a point pilot, i.e., a single impulse located at $(\pilotDelayIdx,\pilotDopplerIdx) \triangleq \left( \frac{\numDelayBin}{2},\frac{\numDopplerBin}{2} \right)$ in the DD domain, is transmitted.
The estimated effective DD domain channel, ${\chEffEstMat}$, can be obtained from the received pilot frame:
\begin{equation}
\label{eq:heff}
{\chEffEstMatEle}[\delayIdx,\DopplerIdx]
=
{\sigRxDD}\left[{\delayIdx}+\tfrac{\numDelayBin}{2},{\DopplerIdx}+\tfrac{{\numDopplerBin}}{2}\right] \cdot
e^{\frac{-\jimg\pi{\DopplerIdx}}{{\numDopplerBin}}},
\end{equation}
where $-\frac{\numDelayBin}{2}\leq {\delayIdx} < \frac{\numDelayBin}{2}$ and $-\frac{{\numDopplerBin}}{2}\leq {\DopplerIdx} < \frac{{\numDopplerBin}}{2}$.
The DD domain channel relationship \eqref{eq:y_channel_DD} can also be described as
\begin{equation}\label{eq:eq_lin_sys}
{\sigRxDDVec} = {\chMatDD}{\sigTxDDVec} + {\chNoiseDDVec},
\end{equation}
with $\sigTxDDVec$, $\sigRxDDVec$, and $\chNoiseDDVec$ the vectorized $\sigTxDDMat$, $\sigRxDDMat$, and $\chNoiseDDMat$.
In this case, the channel matrix ${\chMatDDEst} \in \mathbb{C}^{\numDelayBin {\numDopplerBin} \times \numDelayBin {\numDopplerBin}}$ is obtained from $\chEffEstMat$~\cite[Eq.~(38)]{mohammed2023otfs} using
\begin{equation}\label{eq:Hdd}
\begin{aligned}
&
    {\chEleDDEst}[
        {\delayIdx}'{\numDopplerBin}+{\DopplerIdx}',
        {\delayIdx}{\numDopplerBin}+{\DopplerIdx}
    ]
    =
    \sum_{n,m\in\mathbb{Z}}
    e^{
        \jimg2\pi
        \frac{({\DopplerIdx}'-{\DopplerIdx}-m{\numDopplerBin})({\delayIdx}+n\numDelayBin)}{\numDelayBin {\numDopplerBin}}
    } \\
& \quad
    \times {\chEffEstMatEle}[
        {\delayIdx}'-{\delayIdx}-n\numDelayBin,
        {\DopplerIdx}'-{\DopplerIdx}-m{\numDopplerBin}
    ]
    e^{
            \frac{\jimg2\pi}{{\numDopplerBin}}n{\DopplerIdx}
    },
\end{aligned}
\end{equation}
where ${\delayIdx},{\delayIdx}'\in\delayIdxSet$ and ${\DopplerIdx},{\DopplerIdx}'\in\DopplerIdxSet$.
Note that only one pair of $\left( {\delayIdx}'-{\delayIdx}-n\numDelayBin, {\DopplerIdx}'-{\DopplerIdx}-m{\numDopplerBin} \right)$ falls within the valid indices of $\chEffEstMat$.
The channel $\chMatDDEst$ depicts the pair-wise input/output relationship across the DD domain resource grid points, leading to a large matrix sized ${\numDelayBin}{\numDopplerBin}\times{\numDelayBin}{\numDopplerBin}$.

\subsubsection{\underline{Equalization}}
\label{sub2sec:prelim_eq}

OTFS systems typically rely on LMMSE equalization to jointly mitigate DD interference and noise by inverting the channel matrix, $\chMatDDEst$, which is computationally expensive as the DD grid size $(\numDelayBin,\numDopplerBin)$ increases.
Low-complexity equalizers have been proposed to avoid direct matrix inversion, such as maximum ratio combining (MRC)~\cite{thaj2020mrc} and the iterative conjugate gradient algorithm (CGA)~\cite{mattu2025low}.

\myparatight{Linear minimum mean square error (LMMSE).}
The LMMSE equalizer solves {\eqref{eq:eq_lin_sys}} with direct matrix inversion,
\begin{equation}
\label{eq:lmmse}
{\sigTxDDVecEst}
= \left[ \left({\chMatDDEstHerm}{\chMatDDEst}+{\noiseCovMat}\right)^{-1} {\chMatDDEstHerm} \right] {\sigRxDDVec}
\triangleq {\chMatMMSE} \, {\sigRxDDVec},
\end{equation}
where ${\noiseCovMat} = (\text{SNR}_\text{linear})^{-1} \textbf{I}$ denotes the noise covariance matrix.
It is well-known that LMMSE has a computational complexity of $\bigO({\numDelayBin}^3{\numDopplerBin}^3)$.
In practice, ${\chMatMMSE}$ can also be calculated by solving
\begin{equation}
{\chMatDDEstHerm} = \left( {\chMatDDEstHerm}{\chMatDDEst}+{\noiseCovMat} \right) {\chMatMMSE}
\end{equation}
using factorization techniques such as the Cholesky or lower-upper (LU) decomposition at the same complexity.

\myparatight{Maximum ratio combining (MRC).}
The MRC equalizer~\cite{thaj2020mrc} operates on symbol vectors, where each vector consists of symbols sharing the same delay index across all Doppler indices. Each transmitted symbol vector contributes to multiple received vectors due to DD coupling. MRC extracts these components across delay branches, cancels interference using current estimates, and coherently combines them to maximize SNR. The updated symbol vector is then mapped to the nearest constellation point and refined iteratively in a decision-feedback manner. This process requires hard demodulation at each iteration, introducing additional computational overhead. The overall complexity is $\bigO(\cgaIters\numPath\numDelayBin\numDopplerBin)$, where $\cgaIters$ and $\numPath$ denote the iteration count and the number of paths in $\chEffEstMat$, respectively.

\myparatight{Conjugate gradient algorithm (CGA).}
To avoid explicit matrix inversion on $\chMatDD$, CGA~\cite{mattu2025low} solves {\eqref{eq:lmmse}} in an iterative manner.
Starting from an initial estimate of $\sigTxDDVecEst = \mathbf{0}$, CGA updates the solution using residual-based search directions in each step and terminates when the residual norm falls below a predefined threshold or a maximum number of iterations is reached.
Since each iteration in CGA requires only MVMs~\cite[Algo. 1]{mattu2025low}, it has a computational complexity of $\bigO(\cgaIters\numPath\numDelayBin\numDopplerBin)$ but with lower coefficient compared to MRC.

\subsubsection{\underline{Demodulation}}

We consider hard demodulation given by
\begin{equation}
\label{eq:demod}
\sigTxDDVecEstDemodEle = \arg\min_{s \in \mathcal{S}} \|{\sigTxDDVecEle} - s\|^2,
\quad
\forall {\sigTxDDVecEle} \in {\sigTxDDVec},
\end{equation}
and demap $\sigTxDDVecEstDemodEle$ into their corresponding binary sequences, from which the BER can be obtained.


\begin{figure}[!t]
    \centering
    \vspace{-1.5mm}
    \includegraphics[width=0.98\columnwidth]{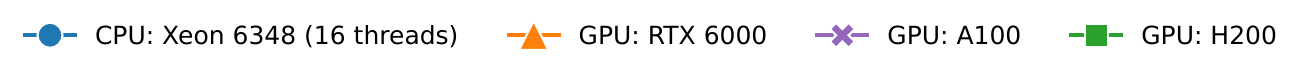}\vspace{-4mm}
    \subfloat[Matrix Inversion]{
    \includegraphics[width=0.45\columnwidth]{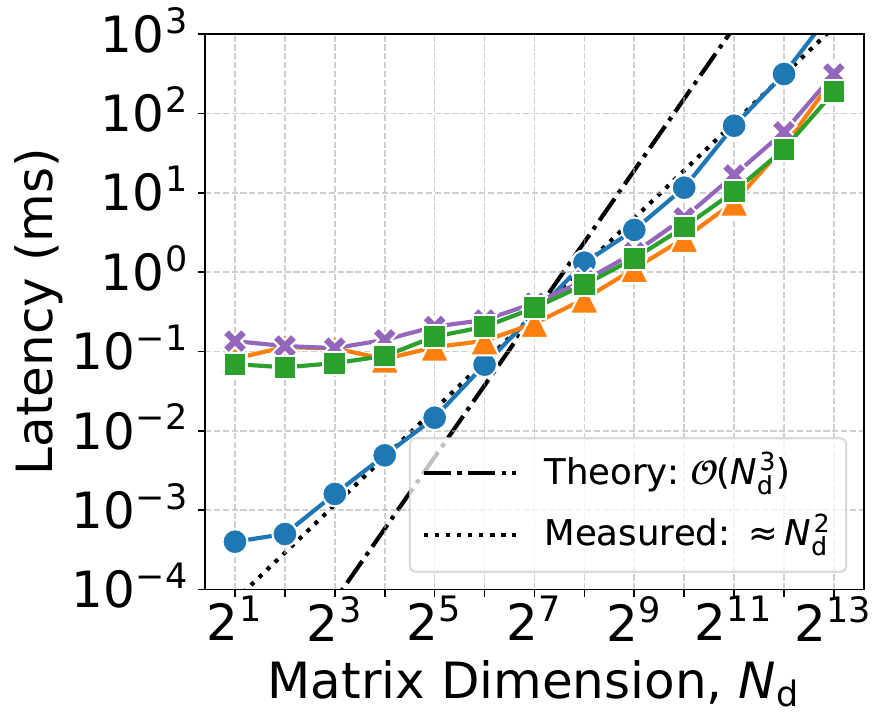}
    \label{subfig:eval_time_unit_minv_plat_cpu16c}}
    \subfloat[MVM]{
    \includegraphics[width=0.45\columnwidth]{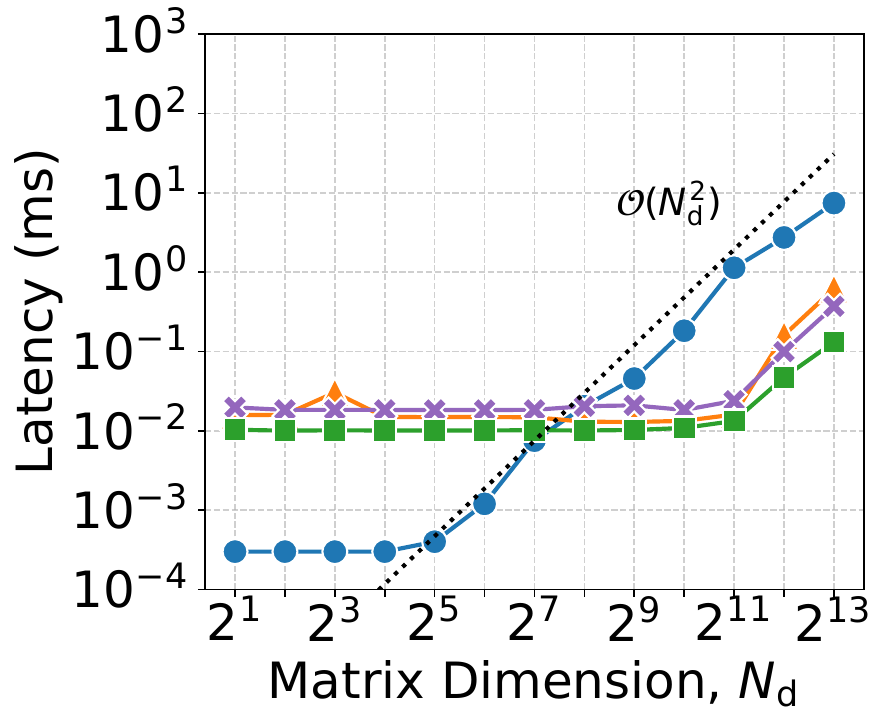}
    \label{subfig:eval_time_unit_mvm_plat_cpu16c}}
    \caption{Measured compute latency and scalability on $\textbf{A} \in \mathbb{C}^{{N_\text{d}}\times{N_\text{d}}}$ across CPUand GPU platforms, including (a) matrix inversion, and (b) matrix-vector multiplication (MVM), with varying matrix dimension $N_\text{d}$.
    GPU-based matrix operations achieve lower latency and better scalability for ${N_\text{d}} > 256$.
    }
    \label{fig:eval_time_unit_mat_op_plat_cpu16c}
    \vspace{-1.5mm}
\end{figure}

\subsection{Scalability of Matrix Operations in Practical Systems}
\label{subsec:scalablity_matop}

While low-complexity algorithms, such as MRC~\cite{thaj2020mrc} and CGA~\cite{mattu2025low} equalizers, are a key to enabling real-time and scalable OTFS signal processing, the available computing resources and their varying capability are often not considered in prior work.
Specifically, the arithmetic operations may exhibit scalability that differs from theory due to hardware specifications and implementations.

To understand how real systems scale with respect to matrix dimensions, we profile the execution latency of matrix operations using a highly optimized BLAS (Basic Linear Algebra Subprograms) library on both CPU and GPU platforms.
Fig.~\ref{fig:eval_time_unit_mat_op_plat_cpu16c} illustrates the execution latency of two operations with varying matrix dimensions, ${N_\text{d}}$:
(\emph{i}) Matrix inversion of $\mathbf{A} \in \mathbb{C}^{{N_\text{d}} \times {N_\text{d}}}$, $\textbf{A}^{-1}$, and
(\emph{ii}) Matrix-vector multiplication (MVM) of $\textbf{y} = \textbf{A} \textbf{x}$ with $\mathbf{A} \in \mathbb{C}^{{N_\text{d}} \times {N_\text{d}}}$ and $\textbf{x}, \textbf{y} \in \mathbb{C}^{N_\text{d}}$.
We consider two compute platforms (for more details see Section~\ref{sec:impl}):
(\emph{i}) server-grade Intel CPU (Xeon 6348) using Intel Math Kernel Library (MKL), and
(\emph{ii}) different NVIDIA GPUs, including RTX 6000 Ada, A100, and H200, using cuBLAS.
We configure MKL to use 16 threads to enable parallel processing on the CPU, and each profiling is executed 10K times

Fig.~\ref{fig:eval_time_unit_mat_op_plat_cpu16c}\subref{subfig:eval_time_unit_minv_plat_cpu16c} shows the mean latency of matrix inversion on CPU/GPU platforms with varying ${N_\text{d}}$.
The measured complexity of inverting $\textbf{A} \in \mathbb{C}^{{N_\text{d}}\times{N_\text{d}}}$ scales in $\bigO({N_\text{d}}^2)$ for both CPU and GPU, with the GPU having a larger overhead when ${N_\text{d}} < 128$, despite the theoretical complexity of $\bigO({N_\text{d}}^3)$.
The result suggests that the processing pipeline design should consider practical scalability within the computing system.
Fig.~\ref{fig:eval_time_unit_mat_op_plat_cpu16c}\subref{subfig:eval_time_unit_mvm_plat_cpu16c} shows the mean latency of MVM on CPU/GPU platforms with varying ${N_\text{d}}$.
The measurements demonstrates that GPU-based processing maintains nearly $\bigO({N_\text{d}})$ scalability over MVM for ${N_\text{d}} \leq 1024$, whereas the CPU latency turns upwards at ${N_\text{d}} = 32$.
On the other hand, GPU has a larger constant overhead than CPU in their flat regions, leading to an intersection around $32 < {N_\text{d}} < 64$.
Even at moderate bandwidths in modern wireless systems, the resulting problem size already falls into an operating region where GPU-based MVM becomes advantageous; for example, $\numDelayBin = 1024$ corresponds to a {30.72}\thinspace{MHz} bandwidth with $\freqSpacing = 30$\thinspace{kHz}.

The latency comparison between matrix inversion and MVM further supports using low-complexity algorithms (e.g., MRC and CGA) that rely on iterative MVM to avoid explicit matrix inversion on GPU platforms.
In particular, MVM is at least one order of magnitude faster than matrix inversion, and up to two orders of magnitude faster with sufficiently large matrices (e.g., ${N_\text{d}} \geq 256$).

\section{System Design}
\label{sec:design}

\begin{table*}[!t]
\centering
\caption{Arithmetic and implementation insights for the proposed Zak-OTFS receiver processing pipeline.}
\label{tab:otfs_pipeline}
\renewcommand{\arraystretch}{1.15}
\begin{tabular}{lccccc}
\hline
\textbf{OTFS RX Stage} &
\textbf{Formulation} &
\textbf{Primary Insight} &
\textbf{Arithmetic} &
\textbf{Complexity} &
\textbf{Memory} \\
\hline
DZT
& {\eqref{eq:zak}}
& Compact Matrix Operation (Section~\ref{subsec:sys_design_precompute},  {\eqref{eq:zak_gemm}})
& GEMM
& $\bigO(\numDelayBin\numDopplerBin^2)$
& $\bigO(\numDelayBin\numDopplerBin+\numDopplerBin^2)$ \\

$\chEffEstMat$ Calculation
& \eqref{eq:heff}
& Compact Matrix Operation (Section~\ref{subsec:sys_design_precompute}, \eqref{eq:heff_hadamard})
& Hadamard Product
& $\bigO(\numDelayBin\numDopplerBin)$
& $\bigO(\numDelayBin\numDopplerBin)$ \\

$\chMatDDEst$ Construction
& \eqref{eq:Hdd}
& Structured Sparsity (SS) (Section~\ref{subsec:sys_design_sparsity})
& Matrix Construction
& $\bigO(\numPath\numDelayBin\numDopplerBin)$
& $\bigO(\numPath\numDelayBin\numDopplerBin)$ \\

Equalization
& \cite{mattu2025low}
& SS and Compute-Aware CGA (Section~\ref{subsec:sys_design_cga}, Algo.~\ref{alg:dia_sparse_cga})
& MVM
& $\bigO(\numPath\numDelayBin\numDopplerBin)$
& $\bigO(\numPath\numDelayBin\numDopplerBin)$ \\

Demodulation
& \eqref{eq:demod}
& Parallel Decision (Trivial)
& Element-wise Op.
& $\bigO(\numDelayBin\numDopplerBin)$
& $\bigO(\numDelayBin\numDopplerBin)$ \\

\hline
\end{tabular}
\end{table*}

\begin{figure*}[!t]
    \centering
    \vspace{-6mm}
    \subfloat[$\chEffEstMat$ Magnitudes]{
    \includegraphics[width=0.6\columnwidth]{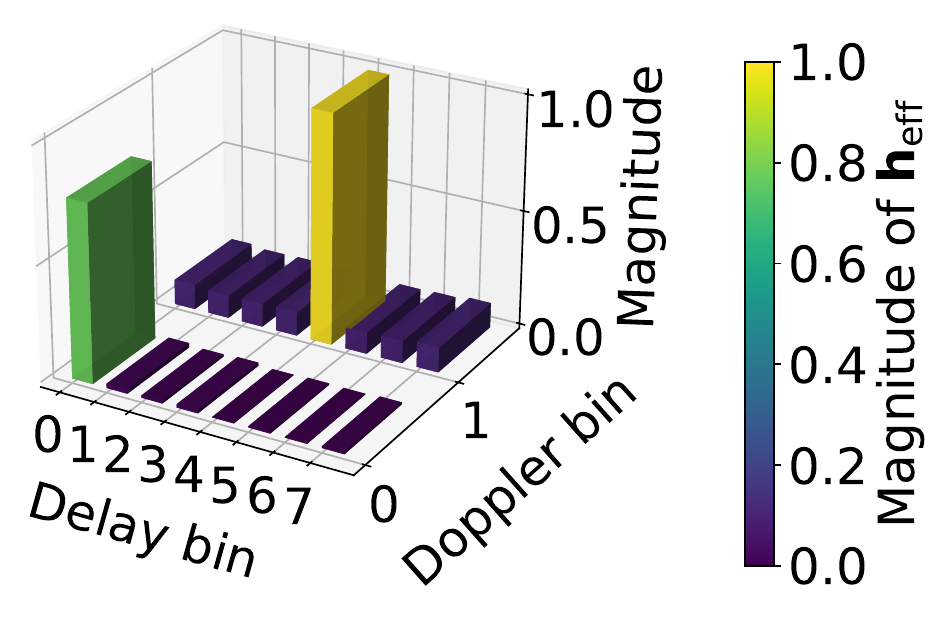}
    \label{subfig:eval_3d_heff}}
    \subfloat[$\chMatDDEst$ Magnitudes]{
    \includegraphics[width=0.52\columnwidth]{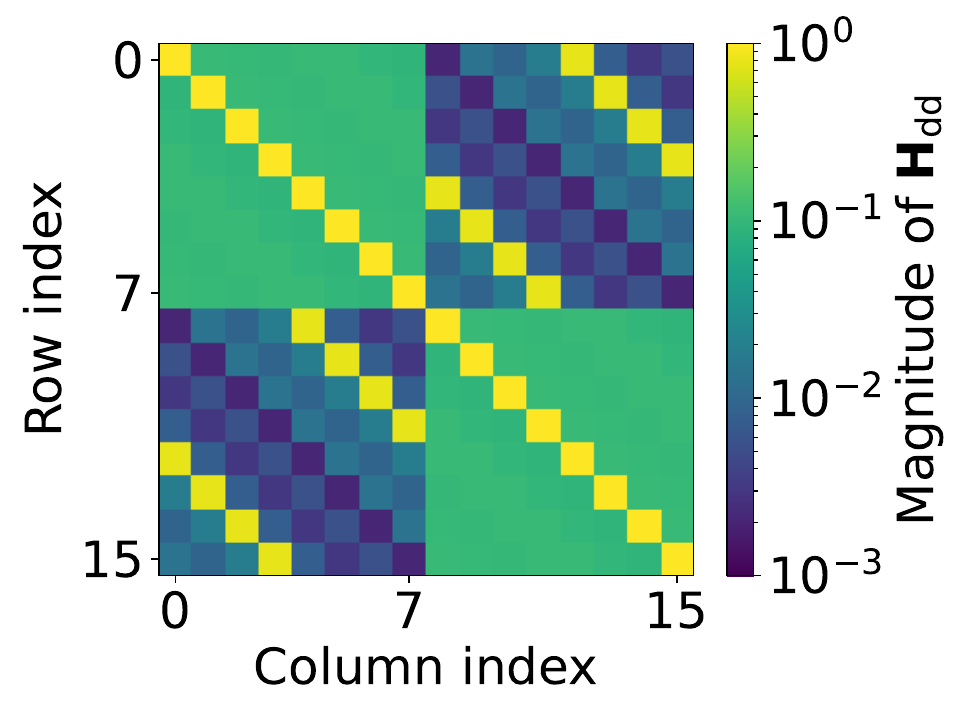}
    \label{subfig:eval_heatmap_Hdd}}
    \subfloat[Binarized $\chEffEstMat$]{
    \includegraphics[width=0.42\columnwidth]{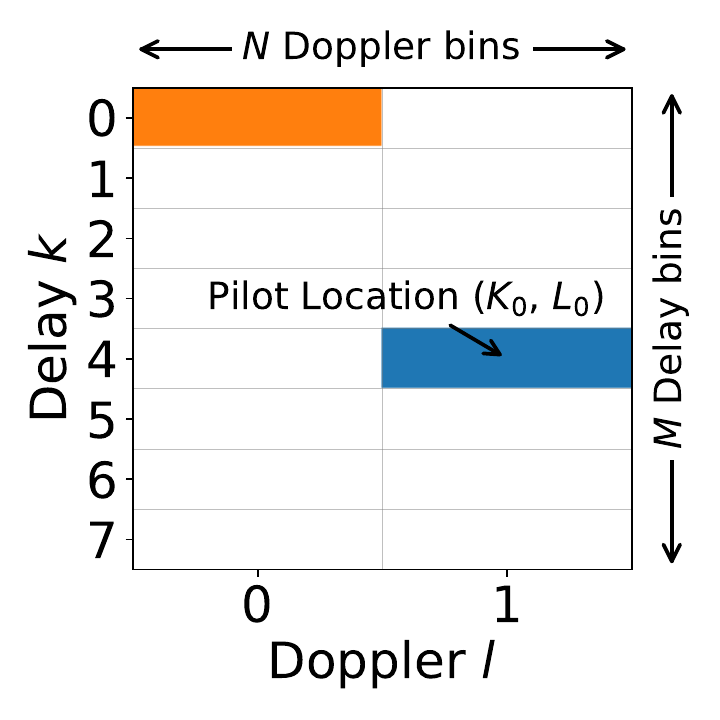}
    \label{subfig:helper_h_eff}}
    \subfloat[Binarized $\chMatDDEst$]{
    \includegraphics[width=0.42\columnwidth]{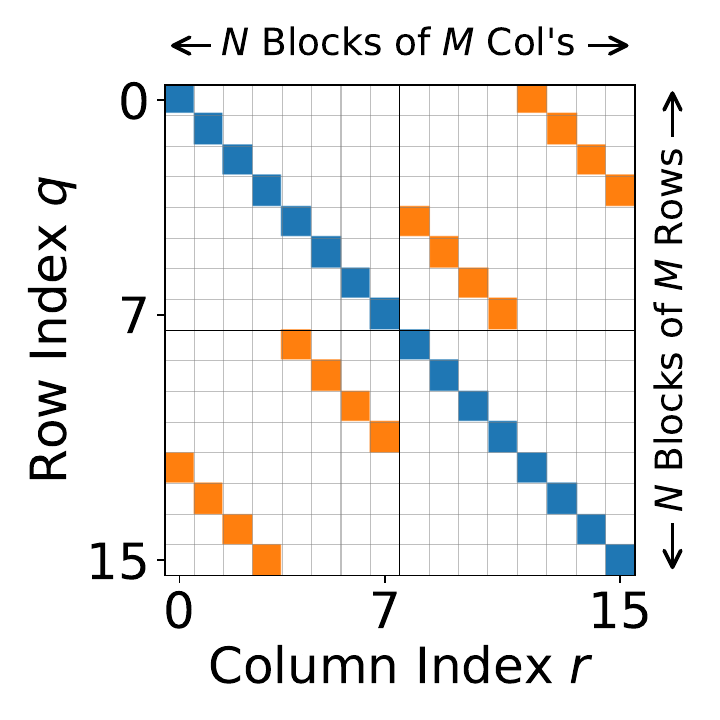}
    \label{subfig:helper_H_dd}}
    \vspace{-1.5mm}
    \caption{
    The $\chEffEstMat$ and $\chMatDDEst$ magnitudes in $(\numDelayBin,\numDopplerBin)=(8,2)$.
    The binarized versions depict the dominant entries using a threshold of $\heffThres = 0.12$, where any entry with magnitude below the threshold is set to zero.
    (a) The significant bins in $\chEffEstMat$ correspond to the dominant channel paths.
    (b) The channel matrix $\chMatDDEst$ is divided into $\numDopplerBin\times \numDopplerBin$ smaller sub-matrices of dimension $\numDelayBin\times \numDelayBin$.
    (c) Thresholding allows isolation of the most significant paths in $\chEffEstMat$, with each path at $(\pathDelayIdx,\pathDopplerIdx)$.
    In this example, two paths are located at (4, 1) in blue and at (0, 0) in orange.
    The blue path is associated with zero delay and zero Doppler shift, while the orange path corresponds to a delay shift of 0.017\thinspace{ms} and a Doppler shift of 15\thinspace{kHz}.
    (d) The resultant patterns in $\chMatDDEst$, where the colors associate the dominant path in binarized $\chEffEstMat$ and $\chMatDDEst$.
    For each row $\chMatDDRowIdx$ in $\chMatDDEst$, the dominant path $\pathIdx$ maps to column $\chMatDDColIdxMappedRowIdx$ as a function of $(\pathDelayIdx,\pathDopplerIdx)$ \eqref{eq:map_chMatDD_col}.
    }
    \label{fig:viz_Hdd_mapping}
    \vspace{-1.5mm}
\end{figure*}

Drawing on these insights, the proposed Zak-OTFS processing pipeline leverages compute-aware optimizations to meet real-time latency requirements while enhancing scalability.
Table~\ref{tab:otfs_pipeline} summarizes the five major DSP stages in the Zak-OTFS receiver which, except for demodulation, require insight into the compute-aware arithmetic design.
Our design accelerates these stages by
\emph{(i)} using compact matrix operation,
\emph{(ii)} leveraging channel sparsity, and
\emph{(iii)} designing a compute-aware algorithm, as detailed next in this section.

\subsection{Compact Matrix Operations}\label{subsec:sys_design_precompute}

In the Zak-OTFS signal processing pipeline (see Section~\ref{subsec:prelim_pipeline}), exponential factors are constantly used in each stage, and are invariant across the frame.
Hence, precomputing those terms offline and applying them on the fly reduces the overhead of repeated runtime computation.
For example, DZT {\eqref{eq:zak}} and estimating $\chEffMat$ {\eqref{eq:heff}} can be decoupled into \emph{offline} computed assistant terms and \emph{runtime} compact matrix operations suitable for highly optimized GPU libraries.

\myparatight{GEMM for DZT.}
Specifically, for DZT, we first construct an auxiliary matrix of FFT kernels $\kernelFftMat \in \mathbb{C}^{\numDopplerBin\times\numDopplerBin}$, given by
\begin{equation}
\label{eqn:zak_exp}
\kernelFftMatEle[\DopplerIdx', \DopplerIdx]
=
e^{-\jimg2\pi \frac{\DopplerIdx' \DopplerIdx}{\numDopplerBin}} \cdot e^{\jimg\pi \DopplerIdx}
=
(-1)^{\DopplerIdx} \cdot e^{-\jimg2\pi \frac{\DopplerIdx' \DopplerIdx}{\numDopplerBin}}.
\end{equation}
Then, we reshape the received time-domain signal vector $\sigRxTdVec \in \mathbb{C}^{\numDelayBin\numDopplerBin}$ into the matrix $\sigRxTdMat \in \mathbb{C}^{{\numDelayBin}\times{\numDopplerBin}}$ in column-major order.
Under these compact matrix representations, DZT can then be performed in one GEMM with complexity $\bigO (\numDelayBin {\numDopplerBin}^2)$ given by
\begin{equation}\label{eq:zak_gemm}
\sigRxDDMat = \sigRxTdMat \kernelFftMat.
\end{equation}
Although the complexity is higher than that of the FFT-based expression of $\bigO(\numDelayBin {\numDopplerBin}\log {\numDopplerBin})$, {\eqref{eq:zak_gemm}} outperforms in speed in the profiled system, as modern GPU arithmetic units are highly optimized for GEMM.

\myparatight{Hadamard product for $\chEffEstMat$.}
Similarly, we create an auxiliary matrix $\twistMat \in \mathbb{C}^{\numDelayBin\times\numDopplerBin}$ to store the coefficients required for the twist convolution in {\eqref{eq:heff}},
\begin{equation}
\label{eq:heff_exp}
    \twistMatEle[\delayIdx, \DopplerIdx] = e^{-\jimg2\pi \frac{\pilotDelayIdx (\DopplerIdx-\pilotDopplerIdx)}{{\numDelayBin}{\numDopplerBin}}},
\end{equation}
and the estimation of $\chEffMat$ can be performed via the Hadamard (element-wise) product given by
\begin{equation}
\label{eq:heff_hadamard}
    \chEffEstMat = \sigRxDDMat \odot \twistMat.
\end{equation}
with a complexity of only $\bigO (\numDelayBin {\numDopplerBin})$.

\subsection{Efficient Matrix Operation via Structured Channel Sparsity}
\label{subsec:sys_design_sparsity}

In OTFS processing, the DD grid dimension of ${\numDelayBin}\times{\numDopplerBin}$, results in a channel matrix $\chMatDD \in \mathbb{C}^{{\numDelayBin}{\numDopplerBin}\times{\numDelayBin}{\numDopplerBin}}$.
This dimension explosion prevents existing OTFS processing pipelines from exploring larger grid sizes, as the massive matrix scales incur prohibitive costs in both computational intensity and memory constraints, including bandwidth and storage requirements.
Note that from {\eqref{eq:heff_model}}, ${\chEffEstMatEle}[{\delayIdx},{\DopplerIdx}]$ is given by the sum of $\numPath$ terms, which is naturally sparse in the DD domain.
Physically, these terms correspond to the discrete delay and Doppler shifts of the $\numPath$ dominant paths of the wireless propagation channel.
Fig.~\ref{fig:viz_Hdd_mapping}\subref{subfig:eval_3d_heff} shows an example visualization of $\chEffEstMat$, which reflects two dominant paths in the channel.
Since $\chMatDDEst$ is obtained from ${\chEffEstMatEle}[{\delayIdx},{\DopplerIdx}]$ using {\eqref{eq:Hdd}} to expand the dimension with no additional information, the resultant $\chMatDDEst$ is both sparse and structured.
By exploiting the structured sparsity of $\chMatDDEst$, we propose an efficient matrix representation that simultaneously minimizes storage footprint and accelerates MVM.

\subsubsection{\underline{Structure of Channel Matrix}}
While the strict block-circulant property of $\chMatDD$ theoretically holds only under integer Doppler shifts, our observations reveal that $\chMatDDEst$ retains a quasi-block-circulant structure even in the more general scenarios.
This is because the entries of $\chMatDDEst$ are deterministically computed by elements in $\chEffEstMat$, allowing for a predictable mapping of dominant paths.
Despite the presence of fractional Doppler leakage, the primary gain distribution of $\chMatDDEst$ consistently follows the indexing pattern of a block-circulant structure.
Consequently, we can exploit this structured sparsity to construct and operate $\chMatDDEst$ in low complexity.

Fig.~\ref{fig:viz_Hdd_mapping}\subref{subfig:eval_heatmap_Hdd} demonstrates how $\chMatDDEst$ exhibit a block-wise diagonal-like pattern even if the corresponding $\chEffEstMat$ in Fig.~\ref{fig:viz_Hdd_mapping}\subref{subfig:eval_3d_heff} is irregular.
Each physical propagation path in the channel induces an independent pattern in $\chMatDDEst$ with circular shift in both row and column directions.
For example, each dominant path (distinguished by colors) located in Fig.~\ref{fig:viz_Hdd_mapping}\subref{subfig:helper_h_eff} leads to a pattern in Fig.~\ref{fig:viz_Hdd_mapping}\subref{subfig:helper_H_dd}.
This block-circulant structure yields $\numDopplerBin\times\numDopplerBin$ blocks, each of size $\numDelayBin\times\numDelayBin$.
Within each block, successive rows are cyclic shifts of one another with linearly increasing offsets; a similar pattern holds column-wise.

\subsubsection{\underline{Exploiting Structured Sparsity}}
\label{sub2sec:ss_mvm}
Note that the $\pathIdx$-th active path corresponding to a pulse at $\chEffMatEle[\pathDelayIdx,\pathDopplerIdx]$ induces a circular shift of the transmitted symbol grid in $\chMatDD$.
We use ${\chMatDDColIdxMapped{\chMatDDRowIdx}}$ to denote the column index of a significant value in the $\chMatDDRowIdx$-the row of $\chMatDDEst$ induced by the dominant path $\pathIdx$.
Hence, each row $\chMatDDRowIdx$ exists $\numPath$ significant values where each is mapped to $\pathIdx \in \pathIdxSet = \{1,2,\ldots,{\numPath}\}$.
With the knowledge of ${\chMatDDColIdxMapped{\chMatDDRowIdx}}$ for $\chMatDDEst$ (to be discussed in Section~\ref{sub2sec:ss_hdd_const}), the MVM using $\chMatDDEst \in \mathbb{C}^{{\numDelayBin}{\numDopplerBin}\times{\numDelayBin}{\numDopplerBin}}$ and a vector $\mathbf{v} \in \mathbb{C}^{{\numDelayBin}{\numDopplerBin}}$
\begin{equation}
    \mathbf{u} = {\chMatDDEst}{\mathbf{v}},
    \quad
    \mathbf{u} \in \mathbb{C}^{{\numDelayBin}{\numDopplerBin}}
\end{equation}
can be approximated via dominant terms by
\begin{equation}
\label{eq:sparse_mvm}
u[{\chMatDDRowIdx}]
=
\sum_{{\chMatDDColIdx}=0}^{{\numDelayBin}{\numDopplerBin}-1}
    {\chEleDDEst}[{\chMatDDRowIdx},{\chMatDDColIdx}]
    \cdot v  [{\chMatDDColIdx}]
\approx
\sum_{{\pathIdx}=1}^{\numPath}
    {\sparseCoe}_{{\pathIdx},{\chMatDDRowIdx}} \cdot
    v[{\chMatDDColIdxMapped{\chMatDDRowIdx}}],\ \forall \chMatDDRowIdx \in {\gridFlatIdxSet},
\end{equation}
where ${\gridFlatIdxSet} = \{0,1,\ldots,{\numDelayBin}{\numDopplerBin}-1\}$ and coefficients are given by
\begin{equation}
\label{eq:D_definition}
{\sparseCoe}_{{\pathIdx},{\chMatDDRowIdx}}
\defn
{\chEleDDEst}[{\chMatDDRowIdx},{\chMatDDColIdxMapped{\chMatDDRowIdx}}].
\end{equation}
Hence, instead of summing over all $\numDelayBin\numDopplerBin$ columns, one only needs to sum over the $\numPath$ path-induced dominant terms for each row (e.g., the colored entries shown in Fig.~\ref{fig:viz_Hdd_mapping}\subref{subfig:helper_H_dd}).
The arithmetic complexity is thus reduced from $\bigO(\numDelayBin^2\numDopplerBin^2)$ to $\bigO(\numPath\numDelayBin\numDopplerBin)$, which is highly advantageous for $\numPath \ll \numDelayBin\numDopplerBin$.
In the example shown in Fig.~\ref{fig:viz_Hdd_mapping}, the number of multiplications is reduced by a factor of 8$\times$ (from 256 to 32).

Similarly, $\chMatDDEstHerm$ requires an inverse mapping that identifies the row index corresponding to a given output column $\chMatDDColIdx$ and path $\pathIdx$.
Multiplying $\chMatDDEstHerm \in \mathbb{C}^{{\numDelayBin}{\numDopplerBin}\times{\numDelayBin}{\numDopplerBin}}$ by a vector $\mathbf{v} \in \mathbb{C}^{{\numDelayBin}{\numDopplerBin}}$ can be written as
\begin{equation}\label{eq:sparse_herm_mvm}
u[{\chMatDDColIdx}]
\approx
\sum_{\pathIdx=1}^{\numPath}
    \sparseCoe^{\ast}_{\pathIdx,\chMatDDRowIdxMappedColIdx} \cdot
    v[{\chMatDDRowIdxMappedColIdx}],
\quad
\forall \chMatDDColIdx \in {\gridFlatIdxSet}.
\end{equation}

Moreover, the reduced MVMs in {\eqref{eq:sparse_mvm}} and {\eqref{eq:sparse_herm_mvm}} have the gather-multiply-reduce computational pattern:
\emph{(i)} \emph{gather} the required input vector entries $v[\cdot]$ using known indices $\chMatDDColIdxMappedRowIdx$ and ${\chMatDDRowIdxMappedColIdx}$;
\emph{(ii)} \emph{multiply} each gathered vector entry by the corresponding path coefficient $\coePathRowIdx$ 
or $\sparseCoe^{\ast}_{\pathIdx,\chMatDDRowIdxMappedColIdx}$
;
\emph{(iii)} \emph{reduce} across the path dimension $\pathIdx$.
This structure avoids irregular sparse traversal and eliminates the need for atomic updates, making it particularly suitable for efficient GPU execution.

\subsubsection{\underline{Efficient Data Structure}}
Section~\ref{sub2sec:ss_mvm} reveals that to perform MVM using $\chMatDDEst$ {\eqref{eq:sparse_mvm}} 
or $\chMatDDEstHerm$ {\eqref{eq:sparse_herm_mvm}}
, storing the entire ${\numDelayBin}{\numDopplerBin}\times{\numDelayBin}{\numDopplerBin}$ dense matrix in unnecessary.
Instead, $\chMatDDEst$ 
and $\chMatDDEstHerm$ 
can be respectively stored in a structured-sparse form as
\begin{equation}\label{eq:ds_repr}
\left\{
    \coePathRowIdx, \chMatDDColIdxMappedRowIdx
\right\}_{\pathIdx=1}^{\numPath},\ \forall \chMatDDRowIdx,\
\text{and}\
\left\{
    \sparseCoe^{\ast}_{\pathIdx,\chMatDDRowIdxMappedColIdx}, \chMatDDRowIdxMappedColIdx
\right\}_{\pathIdx=1}^{\numPath},\ \forall \chMatDDColIdx.
\end{equation}
This structured-sparsity representation requires memory storage with a size of $\bigO({\numPath}{\numDelayBin}{\numDopplerBin})$ elements, which is much smaller than two dense matrices in size $\bigO({\numDelayBin}{\numDopplerBin} \times {\numDelayBin}{\numDopplerBin})$ as ${\numPath} \ll {\numDelayBin}{\numDopplerBin}$.

\subsubsection{\underline{Constructing $\chMatDD$ in Structured-Sparse Representation}}
\label{sub2sec:ss_hdd_const}

We now describe how to construct the structured-sparse (SS) $\chMatDDEst$ in {\eqref{eq:ds_repr}} directly from the channel response $\chEffEstMat$ without explicitly forming the dense matrix in $\mathbb{C}^{{\numDelayBin}{\numDopplerBin}\times {\numDelayBin}{\numDopplerBin}}$.
Since the construction of $\chMatDDEst$ is a 2D circular shift with phase rotation, the position, magnitude, and phase shift of the dominant elements in $\chMatDDEst$ can be computed analytically.

\myparatight{Dominant path determination.}
In typical wireless channels, the number of dominant propagation paths is small, hence $\chEffMat$ is sparse.
We identify the active propagation paths by thresholding the channel magnitude:
\begin{equation}\label{eq:heff_thres}
\chPathThresSet
=
\left\{
    (\delayIdx,\DopplerIdx)
    \;:\;
    \left|\chEffMatEle[\delayIdx,\DopplerIdx]\right| > \heffThres
\right\},
\end{equation}
where $\heffThres$ is a small threshold used to suppress noise.
We represents the delay-bin and Doppler-bin location of the $p$-th path in the DD grid as $({\pathDelayIdx},{\pathDopplerIdx})$ with ${\pathIdx} \in {\pathIdxSet}$, and the corresponding complex channel gains are defined as
\begin{equation}
{\chPathGain}
\defn
{\chEffMatEle}[{\pathDelayIdx},{\pathDopplerIdx}].
\end{equation}

\myparatight{Analytical indices mapping.}
Eq.~{\eqref{eq:Hdd}} depicts the construction of $\chMatDDEst$ from $\chEffEstMat$, and a pulse at $\chEffMatEle[\pathDelayIdx,\pathDopplerIdx]$ caused by a dominant path $\pathIdx$ induces a circular shift of the transmitted symbol grid in $\chMatDD$.
Consequently, for row $\chMatDDRowIdx$ of $\chMatDDEst$, the $\pathIdx$-th active path contributes \emph{only one} dominant matrix entry at column ${\chMatDDColIdxMapped{\chMatDDRowIdx}}$ based on ($\pathDelayIdx,\pathDopplerIdx$).

We index the $\chMatDDRowIdx$-th row in $\chMatDDEst$ by a pair of DD coordinates $({\delayIdx},{\DopplerIdx})$ as $\chMatDDRowIdx = {\DopplerIdx}{\numDelayBin} + {\delayIdx},\ {\delayIdx} \in \delayIdxSet,\ {\DopplerIdx} \in \DopplerIdxSet$.
Equivalently, 
${\delayIdx}_{\chMatDDRowIdx} = \mymod{\chMatDDRowIdx}{\numDelayBin}$
and
${\DopplerIdx}_{\chMatDDRowIdx} = \left\lfloor {\chMatDDRowIdx}/{\numDelayBin} \right\rfloor$
for any row index ${\chMatDDRowIdx} \in {\gridFlatIdxSet}$.
We denote the column index $\chMatDDColIdx$ of the dominant element in $\chMatDDEst$ induced by the $\pathIdx$-th dominant element at row $\chMatDDRowIdx$ as $\chMatDDColIdxMapped{\chMatDDRowIdx}$, which can similarly be expressed as ${\chMatDDColIdxMapped{\chMatDDRowIdx}} = {\DopplerIdx}{\numDelayBin}+{\delayIdx}$ with another set of DD coordinates $({\delayIdx},{\DopplerIdx})$ for ${\delayIdx}\in{\delayIdxSet}$ and ${\DopplerIdx}\in{\DopplerIdxSet}$.
Similarly, the column index ${\chMatDDColIdxMapped{\chMatDDRowIdx}}$ can be expressed with the input DD grid coordinate $(\pathDelayIdx(\chMatDDRowIdx),\pathDopplerIdx(\chMatDDRowIdx))$ as 
\begin{equation}\label{eq:forward_mapping}
{\chMatDDColIdxMapped{\chMatDDRowIdx}}
=
{\pathDopplerIdx(\chMatDDRowIdx)}{\numDelayBin} + {\pathDelayIdx(\chMatDDRowIdx)},
\end{equation}
the output coordinate $({\DopplerIdx}_{\chMatDDRowIdx},{\delayIdx}_{\chMatDDRowIdx})$ for
$
\chMatDDRowIdx
=
{\DopplerIdx}_{\chMatDDRowIdx}{\numDelayBin} + {\delayIdx}_{\chMatDDRowIdx}
$
contributes to the coordinate transform as
\begin{equation}
\pathDelayIdx(\chMatDDRowIdx) = \mymod{\pilotDelayIdx + {\delayIdx}_{\chMatDDRowIdx} - \pathDelayIdx}{\numDelayBin},
\quad
\pathDopplerIdx(\chMatDDRowIdx) = \mymod{\pilotDopplerIdx + {\DopplerIdx}_{\chMatDDRowIdx} - \pathDopplerIdx}{\numDopplerBin},
\end{equation}
where $(\pilotDelayIdx,\pilotDopplerIdx)$ denotes the pilot reference location in the DD grid.
Note that the delay bin offset $\chMatDDColDelayBinOffset$ and the Doppler bin offset $\chMatDDColDopplerBinOffset$ only depend on the pulse location $(\pathDelayIdx,\pathDopplerIdx)$ of $\pathIdx$-th path and the pilot reference location $(\pilotDelayIdx,\pilotDopplerIdx)$ in $\chEffEstMat$:
\begin{equation}
{\chMatDDColDelayBinOffset} \defn \pilotDelayIdx - \pathDelayIdx, \quad
{\chMatDDColDopplerBinOffset} \defn \pilotDopplerIdx - \pathDopplerIdx,
\end{equation}
so that
\begin{equation}\label{eq:map_chMatDD_col}
{\chMatDDColIdxMapped{\chMatDDRowIdx}}
=
{\mymod{{\DopplerIdx}_{\chMatDDRowIdx} + {\chMatDDColDopplerBinOffset}}{\numDopplerBin}} \cdot {\numDelayBin} +
\mymod{{\delayIdx}_{\chMatDDRowIdx} + {\chMatDDColDelayBinOffset}}{\numDelayBin}.
\end{equation}
Fig.~\ref{fig:viz_Hdd_mapping} provides an example with
grid dimension $(\numDelayBin, \numDopplerBin) = (8, 2)$ and
pilot location $(\pilotDelayIdx,\pilotDopplerIdx)=(4,1)$.
The channel has two paths located
$({\delayIdx}_{\pathIdx=0},{\DopplerIdx}_{\pathIdx=0}) = (4,1)$ and
$({\delayIdx}_{\pathIdx=1},{\DopplerIdx}_{\pathIdx=1}) = (0,0)$.
In the row $\chMatDDRowIdx=7=0\numDelayBin+7$, we have
${\DopplerIdx}_{\chMatDDRowIdx=7}=0$ and
${\delayIdx}_{\chMatDDRowIdx=7}=7$.
Hence,
$
{\chMatDDColIdx_{\pathIdx=0}(\chMatDDRowIdx=7)}
=
    \mymod{
        {\DopplerIdx}_{\chMatDDRowIdx=7} +
        {\pilotDopplerIdx-{\DopplerIdx}_{\pathIdx=0}}}{\numDopplerBin}
    \cdot
    {\numDelayBin} +
    \mymod{
        {\delayIdx}_{\chMatDDRowIdx=7} +
        {\pilotDelayIdx - {\delayIdx}_{\pathIdx=0}}}{\numDelayBin}
=
    \mymod{
        0 +
        {1-1}}{2}
    \cdot
    {8} +
    \mymod{
        7 +
        {4 - 4}}{8}
=7.$
and
$
{\chMatDDColIdx_{\pathIdx=1}(\chMatDDRowIdx=7)}
=
    \mymod{
        0 +
        {1-0}}{2}
    \cdot
    {8} +
    \mymod{
        7 +
        {4 - 0}}{8}
=11.$

For MVM using matrix $\chMatDDEstHerm$, we instead need, for each output index $\chMatDDColIdx$, the row index whose forward mapping equals $\chMatDDColIdx$.
We therefore define the inverse mapping $\chMatDDRowIdxMapped{\chMatDDColIdx}$ such that
\begin{equation}
\chMatDDColIdxMapped{\chMatDDRowIdxMapped{\chMatDDColIdx}} = {\chMatDDColIdx}.
\end{equation}
Let
${\delayIdx}_{\chMatDDColIdx} = \mymod{\chMatDDColIdx}{\numDelayBin}$
and
${\DopplerIdx}_{\chMatDDColIdx} = \lfloor {{\chMatDDColIdx}}/{\numDelayBin} \rfloor$
denote the DD coordinates corresponding to index ${\chMatDDColIdx}$.
Furthermore, define the path-dependent offsets
\begin{equation}
{\chMatDDRowDelayBinOffset} \defn \mymod{\chMatDDColIdxMapped{0}}{\numDelayBin} = {\pathDelayIdx(0)},
\quad
{\chMatDDRowDopplerBinOffset} \defn \lfloor \tfrac{\chMatDDColIdxMapped{0}}{\numDelayBin} \rfloor = {\pathDopplerIdx(0)},
\end{equation}
which correspond to the delay and Doppler shifts induced by path $p$.
The inverse mapping is given by
\begin{equation}\label{eq:herm_mapping}
{\chMatDDRowIdxMapped{\chMatDDColIdx}}
=
\mymod{{\DopplerIdx}_{\chMatDDColIdx}-{\chMatDDColDopplerBinOffset}}{\numDopplerBin} \cdot {\numDelayBin}
+
\mymod{{\delayIdx}_{\chMatDDColIdx}-{\chMatDDColDelayBinOffset}}{\numDelayBin}.
\end{equation}

\myparatight{Phase correction and coefficient synthesis.}
In addition to the index mapping and the channel gain, each matrix entry must include the appropriate OTFS phase rotation, i.e., a complex phase compensation factor ${\phCompFactor}$ which contributes to ${\coePathRowIdx}$ as
\begin{equation}
{\phCompFactor} = e^{\jimg\phTwisted},
\quad
{\coePathRowIdx} = {\chPathGain} {\phCompFactor}
\end{equation}
The quasi-periodic phase term $\phTwisted$ for path $\pathIdx$ and row $\chMatDDRowIdx$ in $\chMatDDEst$ comes from 
\begin{equation}
\phTwisted
=
\tfrac{2\pi}{{\numDelayBin}{\numDopplerBin}}
\left[
(\pathDopplerIdx - \pilotDopplerIdx) \rawDelayOffset
+
\DelayOffsetWrapCnt {\DopplerIdx}_{\pathIdx,{\chMatDDRowIdx}} {\numDelayBin}
\right],
\end{equation}
where $\rawDelayOffset$ is the raw delay index offset and $\DelayOffsetWrapCnt$ is the wrap count due to circular shift
\begin{equation}
\rawDelayOffset \defn \pilotDelayIdx + \delayIdx_{\chMatDDRowIdx} - \pathDelayIdx,
\quad
\DelayOffsetWrapCnt \defn \left\lfloor {\rawDelayOffset}/{\numDelayBin} \right\rfloor.
\end{equation}

\subsubsection{\underline{Discussions on Complexity and Memory Requirements}}

The proposed structured-sparse construction avoids explicitly forming the dense channel matrix $\chMatDDEst \in \mathbb{C}^{{\numDelayBin}{\numDopplerBin} \times {\numDelayBin}{\numDopplerBin}}$, which requires storing $({\numDelayBin}{\numDopplerBin})^2$ complex-valued entries, and a dense MVM has a computational complexity of $\bigO(({\numDelayBin}{\numDopplerBin})^2)$.
Instead, following Section~\ref{sub2sec:ss_hdd_const} requires only $\bigO({\numPath}{\numDelayBin}{\numDopplerBin})$ memory and time to construct the matrix in 
the sparse form \eqref{eq:ds_repr}.
Similarly, \eqref{eq:sparse_mvm} 
and \eqref{eq:sparse_herm_mvm} 
have arithmetic complexity $\bigO({\numPath}{\numDelayBin}{\numDopplerBin})$ to perform MVM.
For example, with $(\numDelayBin,\numDopplerBin)=(48,32)$, $\numDelayBin\numDopplerBin=1{,}536$. 
Assuming $\numPath=5$ paths, the dense representation requires $\approx${2.4}\thinspace{M} complex values, whereas the diagonal-sparse representation stores only $7{,}680$, resulting in a $307.2\times$ reduction in memory requirement.
As we scale to larger Zak-OTFS grids, e.g., $(\numDelayBin,\numDopplerBin)=(8192,32)$, storing the dense channel matrix alone would require $8 \times 8192^2 \times 32^2\thinspace\text{Bytes} \approx 549.8\thinspace\text{GB}$ memory per frame, which needs to be updated every {1.07}\thinspace{\msec} and quickly becomes a significant memory bottleneck.

The key system-level insight is, therefore, that the OTFS channel matrix is not merely sparse but \emph{structurally sparse}.
This structure allows the original dense operator to be replaced by a compact path-indexed lookup representation, preserving the low arithmetic complexity of sparse processing while enabling GPU-friendly memory access patterns that are much more regular than those of a generic sparse matrix format.

\subsection{Compute-Aware Conjugate Gradient-based Equalizer}
\label{subsec:sys_design_cga}

\begin{algorithm}[!t]
  \caption{\small Compute-aware Conjugate Gradient Algorithm (CGA) Equalization.}
  \label{alg:dia_sparse_cga}
  \small
  \begin{algorithmic}[1]
    \Require Structured-sparse channel $\chMatDDEst$ and $\chMatDDEstHerm$, received vector $\sigRxDDVec$, noise covariance matrix $\noiseCovMat$, iterations $\cgaIters=10$
    \Ensure Estimated transmit vector ${\sigTxDDVecEst}$

    \State $\mathbf{b} \gets \chMatDDEstHerm\sigRxDDVec$, ${\sigTxDDVecEst} \gets \mathbf{0}$, $\mathbf{c} \gets \mathbf{b}$, $\mathbf{p} \gets \mathbf{b}$, ${\sqResNorm} \gets \lVert \mathbf{c} \rVert_2^2$

    \For{${\cgaIter} = 1$ to $\cgaIters$}
        \State $\chMatDD' \gets \chMatDDEstHerm(\chMatDDEst \mathbf{p})$ \Comment{Break into two MVMs} \label{line:cga_mvm}
        \State $\mathbf{a}_\mathbf{p} \gets \chMatDD' + \lambda \mathbf{p}$ \Comment{$\noiseCovMat = \lambda I$} \label{line:noise_cov_mat}
        \State $\alpha \gets {\sqResNorm} / (\mathbf{p}^\mathrm{H} \mathbf{a}_\mathbf{p})$
        \State ${\sigTxDDVecEst} \gets {\sigTxDDVecEst} + \alpha \mathbf{p}$, $\mathbf{c} \gets \mathbf{c} - \alpha \mathbf{a}_\mathbf{p}$
        \State ${\sqResNorm}' \gets \lVert \mathbf{c} \rVert_2^2$ \Comment{No conditional early termination}\label{line:c_norm_compute}
        \State $\mathbf{p} \gets \mathbf{c} + ({\sqResNorm}' / {\sqResNorm}) \mathbf{p}$, ${\sqResNorm} \gets {\sqResNorm}'$
    \EndFor

    \State \Return ${\sigTxDDVecEst}$
  \end{algorithmic}
\end{algorithm}

Motivated by the practical scalability of the computation system discussed in Section~\ref {subsec:scalablity_matop}, equalizers using MVM instead of matrix inversion are preferred for GPU-based real-time OTFS processing, e.g., CGA proposed in~\cite{mattu2025low} (see also Section~\ref{sub2sec:prelim_eq}).
However, compute-aware adoption is necessary to translate operator-level scalability in Fig.~\ref{fig:eval_time_unit_mat_op_plat_cpu16c} to the system level.
In addition, unlike the original CGA~\cite{mattu2025low}, which performs equalization in the frequency domain, we carry out conjugate gradient-based equalization directly in the DD domain.
First, transforming between domains (e.g., DD to frequency domain and back) incurs additional computational overhead and latency, as it requires multiplications involving dense transform matrices applied to the vectorized DD transmit signal $\sigTxDDVec$.
In addition, as shown in Section~\ref{subsec:sys_design_sparsity}, the effective channel $\chMatDDEst$ exhibits inherent sparsity in the DD domain due to the limited number of propagation paths. Consequently, operating in the DD domain enables us to exploit the channel sparsity while applying CGA efficiently without incurring transformation overhead.

Algo.~\ref{alg:dia_sparse_cga} shows our proposed compute-aware CGA with three key features:
\emph{(i)} substitute of GEMM with equivalent MVM, 
\emph{(ii)} dimension reduction of operator, and
\emph{(iii)} fixed termination based on offline profiling.

\myparatight{Leveraging scalable operators.}
In the original CGA~\cite{mattu2025low}, the Gram matrix ${\chMatDDEstHerm} \cdot {\chMatDDEst}$ is computed once ahead of the iterative processes.
This incurs a complexity of $\bigO(MN^2)$, which can be reduced to $\bigO(P^2MN)$ leveraging the structured sparsity in $\chMatDD$.
On the other hand, our implementation requires only two MVMs (Line~\ref{line:cga_mvm}, Algo.~\ref{alg:dia_sparse_cga}), therefore further reducing the computational complexity to $\bigO(PMN)$.

\myparatight{Dimension reduction of noise covariance.}
Similarly, for the noise covariance matrix $\noiseCovMat = \lambda \cdot \textbf{I}$, computing $\noiseCovMat$ explicitly incurs dimension increase as $\noiseCovMat \in \mathbb{C}^{{\numDelayBin\numDopplerBin}\times{\numDelayBin\numDopplerBin}}$.
If unified noise variance across the DD bins is assumed, $\noiseCovMat$ can be reduced to a scalar $\lambda = 1/(\text{SNR}_\text{linear})$ (Line~\ref{line:noise_cov_mat}, Algo.~\ref{alg:dia_sparse_cga}).


\begin{figure}[!t]
    \centering
    \vspace{-1.5mm}
    \hspace{8mm}\includegraphics[width=0.7\columnwidth]{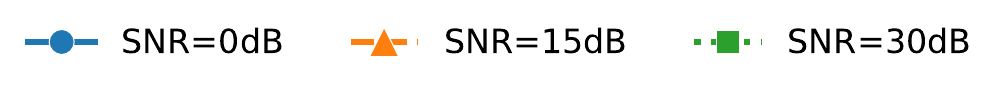}\vspace{-5mm}
    \subfloat[${\sqResNorm}$ vs. CGA Iterations]{
    \includegraphics[width=0.45\columnwidth]{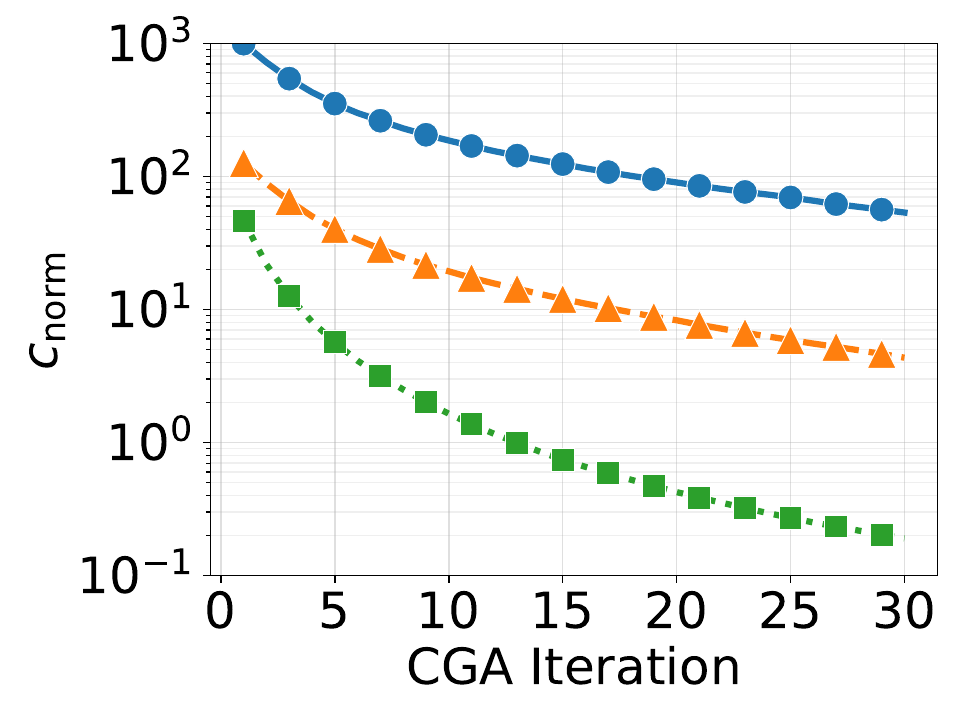}
    \label{subfig:eval_cga_cnorm_vs_iter_M128_N32}}
    \subfloat[BER vs. CGA Iterations]{
    \includegraphics[width=0.45\columnwidth]{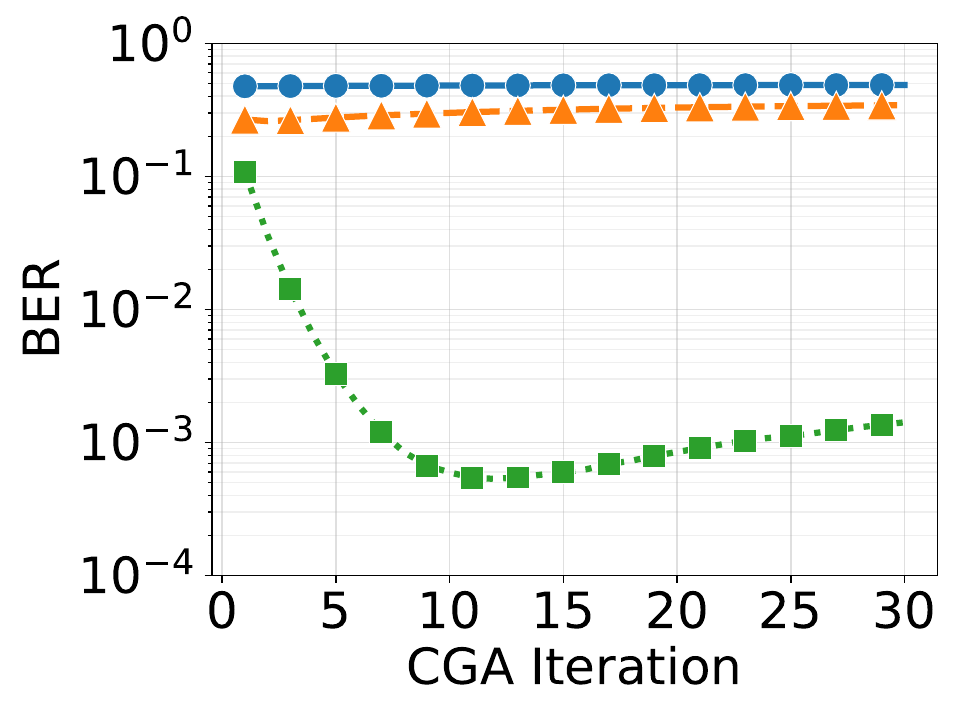}
    \label{subfig:eval_cga_ber_vs_iter_M128_N32}}
    \caption{BER and ${\sqResNorm}$ vs. iterations in CGA for $(\numDelayBin,\numDopplerBin)=(128,32)$, 16QAM.
    Across iterations, ${\sqResNorm}$ exhibits monotonic decay with a diminishing slope, whereas BER behaves more variably, including flat (SNR = 0 dB) and rebound (SNR = 30 dB). In addition, ${\sqResNorm}$ is exponentially proportional to SNR, while BER does not have this behavior.}
    \label{fig:eval_cga_ber_cnorm_vs_iter_M128_N32}
    \vspace{-1.5mm}
\end{figure}

\myparatight{Profiling-based fixed termination.}
While the original CGA benefits from loss-based early termination to reduce iteration counts, this strategy is not well-suited to GPU execution.
This is because implementing early termination requires synchronization across the entire GPU and often involves transferring intermediate results to the host for convergence evaluation, i.e., checking the squared residual norm to stop the iteration when ${\sqResNorm} < \epsilon^2$, with $\epsilon$ being the threshold.
The introduced kernel launch overhead and CPU-GPU synchronization in each iteration can stall the execution pipeline and offset the benefits of reduced arithmetic operations.
Instead, we employ an offline profiling-based fixed iterations for CGA based on the OTFS parameters and link conditions (e.g, DD domain grid size, modulation order, and SNR).

\myparatight{BER-driven iteration selection.}
To determine the iteration budget, we adopt a BER-driven offline profiling approach.
We define the convergence of BER by monitoring the signed relative BER change $r$ between consecutive CGA iterations:
\begin{equation}\label{eq:ber_convergence}
r_{\cgaIter}
=
\frac{\mathrm{BER}_{{\cgaIter}-1} - \mathrm{BER}_{\cgaIter}}{\max(\mathrm{BER}_{\cgaIter-1}, \zeta)},
\end{equation}
where $\zeta > 0$ is a small constant to ensure numerical stability.
The BER is considered converged at the first iteration ${\cgaIter} \geq 2$ such that $r_{\cgaIter} < \eta$, where $\eta$ is a small tolerance.
This criterion captures both cases where the BER begins to rise and where the BER improvement becomes negligible.
We set $\eta = 10^{-2}$ and $\zeta = 10^{-12}$ in all experiments.
The iteration budget is then determined as the maximum convergence iteration across SNRs and grid configurations.

To justify this BER-driven criterion, we examine the relationship between ${\sqResNorm}$ and BER across CGA iterations.
Fig.~\ref{fig:eval_cga_ber_cnorm_vs_iter_M128_N32} plots the ${\sqResNorm}$ values and BER across the CGA iterations under different SNR scenarios.
As expected, ${\sqResNorm}$ decreases monotonically with the iteration count for all SNR levels, reflecting the convergence behavior of the conjugate gradient solver toward the least-squares solution.
However, the BER curves exhibit significantly different trends: it does not necessarily decrease monotonically with iterations even though ${\sqResNorm}$ continues to decrease.
For example, at high SNR (e.g., {30}\thinspace{dB}), the BER first decreases and then slightly increases after reaching a minimum, whereas at lower SNRs (e.g., {0}\thinspace{dB} and {15}\thinspace{dB}), the BER remains largely unchanged across iterations despite the substantial reduction in ${\sqResNorm}$.
Moreover, the same ${\sqResNorm}$ value can correspond to substantially different BER levels under different SNR conditions.
For example, for ${\sqResNorm}$ around $10^{-1}$, BER is $10^{-2}$ under {30}\thinspace{dB} SNR and is larger than $10^{-1}$ under {15}\thinspace{dB} SNR.
These observations indicate that ${\sqResNorm}$ cannot reliably predict BER across operating regimes, making it unsuitable as a universal convergence indicator.

The iteration budget is determined offline using the BER-driven criterion described above, and a fixed-iteration execution model is adopted during runtime.
While condition-based early termination is common in CPU implementations, it is inherently inefficient for the GPU's SIMT architecture due to the costly global synchronization and host-device interactions to decide whether to launch the subsequent demodulation kernel.
Predetermined termination avoids unnecessary runtime monitoring while enabling predictable performance with efficient hardware utilization: executing a fixed number of iterations allows the computation to be organized into a fully GPU-resident pipeline with minimal host intervention.
For example, for a DD domain grid size of $(128, 32)$ with 10 conjugate gradient iterations, eliminating CPU-GPU synchronization reduces the equalization time from {0.65}\thinspace{ms} to {0.54}\thinspace{ms} on average, yielding a 17\% speedup.

\section{Implementation}
\label{sec:impl}

\myparatight{Software implementation and hardware platforms.}
We implement the GPU-based Zak-OTFS processing pipeline in approximately 3,200 lines of Python code.
PyTorch is used as the primary backend for general tensor computation, and OpenAI Triton~\cite{tillet2019triton} is used to develop custom GPU kernels for large-scale matrix operations and performance-critical components, such as MVMs and GEMMs.
We evaluate our implementation of the end-to-end Zak-OTFS receiver processing primarily on an NVIDIA H200 GPU.
We also consider multiple NVIDIA GPU platforms, including Jetson Orin, RTX 6000 Ada, and A100, to provide a comprehensive performance comparison.
To understand the impact of the hardware architecture, we also implement the same Zak-OTFS processing pipeline in C++ using the Armadillo library~\cite{sanderson2025armadillo} and evaluate it on a Dell PowerEdge R750 server, equipped with two Intel Xeon Gold 6348 processors, each with 28 physical cores (56 threads) running at {2.6}\thinspace{GHz} base frequency.
Table~\ref{tab:hardware} provides the detailed specifications of the considered hardware platforms.

\begin{table}[!t]
\centering
\caption{Hardware Platforms Used for Evaluation}
\label{tab:hardware}
\renewcommand{\arraystretch}{1.15}
\begin{tabular}{|c|c|c|}
\hline
\textbf{Type} & \textbf{Platform} & \textbf{Memory} \\
\hline
CPU & Intel Xeon Gold 6348 & DDR4 {256}\thinspace{GB} \\
\hline
\multirow{4}{*}{GPU}
& NVIDIA Jetson AGX Orin & LPDDR5 {64}\thinspace{GB} \\
\cline{2-3}
& NVIDIA RTX 6000 Ada & GDDR6 {48}\thinspace{GB} \\
\cline{2-3}
& NVIDIA A100 & HBM2e {40}\thinspace{GB} \\
\cline{2-3}
& NVIDIA H200 & HBM3e {141}\thinspace{GB} \\
\hline
\end{tabular}
\end{table}


\begin{table}[!t]
    \centering
    \caption{Power-delay Profile of Veh-A Channel Model}
    \renewcommand{\arraystretch}{1.15}
    \begin{tabular}{|c|c c c c c c|}
    \hline
       \textbf{Path Index}, $p$  & 1 & 2 & 3 & 4 & 5 & 6 \\
       \hline
       \textbf{Delay} $\delay_{p}$ (\usec)  & 0.00 & 0.31 & 0.71 & 1.09 & 1.73 & 2.51 \\
       \hline
       \textbf{Relative Power (dB)} & $0$ & $-1$ & $-9$ & $-10$ & $-15$ & $-20$ \\
       \hline
    \end{tabular}
    \label{tab:vehAProf}
\end{table}

\myparatight{Channel model.}
We verify the proposed system using the ITU Vehicular-A (Veh-A) multipath channel model~\cite{itur_m1225} to model a realistic high-mobility propagation environment, which was also adopted in a number of previous works~\cite{mattu2025low, zheng2025zak, mattu2025differential, mehrotra2025zak, dabak2024zak}.
Table~\ref{tab:vehAProf} summarizes the Veh-A channel profile, including six discrete paths with excess delays and their average power levels.
The Doppler shift of the $\pathIdx$-th path is drawn as $\nu_\pathIdx = \chMaxDopplerFreq \cos(2\pi U_\pathIdx)$, where $U_\pathIdx \sim \mathcal{U}(0,1)$ (uniform distribution), with $\chMaxDopplerFreq$ being the maximum Doppler frequency determined by the carrier frequency and vehicle speed.
Note that both the delay and Doppler are fractional.
Additive white Gaussian noise (AWGN) is then applied to the received signal with the noise standard deviation set as $\sigma = \sqrt{\rho_s / \gamma}$, where $\rho_s$ is the mean signal power and $\gamma$ is the SNR in linear scale.

\section{Evaluation}
\label{sec:eval}

In this section, we present a comprehensive evaluation of the proposed Zak-OTFS processing system in terms of real-time capability, latency distribution, scalability, and BER with varying compute platforms, equalizers, and data structures.

\subsection{Experiment Setup and Evaluation Metrics}
\label{subsec:exp_setup_eval_metrics}

\myparatight{Experiment setup.}
We consider a Zak-OTFS system with $\numDelayBin$ delay bins and $\numDopplerBin = 32$ Doppler bins, a frequency spacing of $\freqSpacing = \numDopplerBin\cdot\DopplerRes = 30$\thinspace{kHz}, corresponding to a Doppler resolution of $\DopplerRes=937.50$\thinspace{Hz} and frame duration of $\frameTime=1.07$\thinspace{\msec} {\eqref{eq:bandwidth_frametime}}.
The delay resolution $\delayRes$ depends on the bandwidth $\bandwidth = \numDelayBin \freqSpacing$, given by {\eqref{eq:delay_Doppler_res}}--{\eqref{eq:bandwidth_frametime}}.
We set the value of the sampling rate equal to that of $\bandwidth$.
Quadrature amplitude modulation (QAM) is employed, with QPSK and 16-QAM considered.
The channel coherence time $\coherenceTime$ is inversely proportional to the Doppler spread, i.e., $\coherenceTime \approx 0.423/\chMaxDopplerFreq$.
As the Doppler spread increases, the channel varies more rapidly within a frame, reducing temporal correlation and making channel estimation more challenging. 
We set $\chMaxDopplerFreq=100$\thinspace{Hz} in the Veh-A channel simulation for all latency-related experiments, unless otherwise specified. 
This corresponds to a coherence time on the order of $\coherenceTime \approx 4.23$\thinspace{\msec}, which is significantly larger than the OTFS frame duration $\frameTime=1.07$\thinspace{\msec}. This ensures that the channel remains approximately constant over a frame, allowing us to focus on computational performance and system latency without being dominated by channel aging effects.
The estimated (sparse) DD domain channel matrix $\chMatDDEst$ {\eqref{eq:heff_thres} is constructed using a threshold $\heffThres=0.08$.
The equalizers, including CGA (Algo.~\ref{alg:dia_sparse_cga}) and MRC~\cite{thaj2020mrc}, are executed for $\cgaIters=10$ iterations for a fair comparison, and the effectiveness of this iteration count is discussed in Section~\ref{subsec:cga_ber_converge}.

\myparatight{Real-time processing latency deadline.}
A high deadline satisfaction rate is necessary for reliable communication services~\cite{ding2020agora}, as deadline violations can lead to packet drop and/or backlog accumulation~\cite{tung2026rise}.
Hence, we define real-time processing capability as meeting 99.9\% latency within the deadline, consistent with prior literature on 5G vRAN processing for OFDM signals~\cite{qi2024savannah}.
Note that in some evaluations, we report median latency because the prohibitive execution latencies of the baseline methods prevent comprehensive data collection.
We define the processing deadline of a packet of one pilot and one data frame as two frame durations ($2\frameTime$): the combined pilot and data frame must be processed before the next packet is fully received.
This setup assumes the worst-case scenario that the channel is estimated every other frame, whereas in the real-world scenario, the channel coherence time is longer.
Unless otherwise specified, we consider pilot/data frame duration of $\frameTime=1.067$\thinspace{\msec} with $\numDopplerBin=32$ and $\freqSpacing=30$\thinspace{kHz} so that the packet deadline is $2\frameTime=2.13$\thinspace{\msec}.

\myparatight{BER calculation and throughput.}
We consider hard demodulation and the demodulated bits are compared with the ground truth to calculate frame-level BER.
The BER is averaged across multiple frames.
The throughput, ${\tput}$, is given by
\begin{equation}
\tput
=
\frac{1}{2} \cdot \bandwidth \cdot b_{\text{mod}} \cdot (1 - \mathrm{BER}),
\end{equation}
where $\bandwidth$ equals the sample rate with a factor of $\frac{1}{2}$ due to the interleaving pilot/data frame arrangement, $b_{\text{mod}}$ is the number of bits per DD domain symbol based on the modulation scheme.
With QPSK modulation, our system achieves a data rate of 491.44\thinspace{Mbps} with $\numDelayBin=16{,}384$, $b_{\text{mod}}=2$, and $\mathrm{BER} = 0.015\%$ at {25}\thinspace{dB} SNR.
With 16QAM modulation, our system further pushed the data rate over {906.52}\thinspace{Mbps} with $b_{\text{mod}}=4$ and $\mathrm{BER} = 7.78\%$ under the same configuration.

\myparatight{Baseline equalizers.}
We consider the LMMSE and MRC equalizers introduced in Section~\ref{sub2sec:prelim_eq}, and refer to the proposed compute-aware CGA in Section~\ref{subsec:sys_design_cga} as CGA in the evaluation section.
When the channel matrix $\chMatDDEst$ is using the structured sparsity (SS) representation {\eqref{eq:ds_repr}} in Section~\ref{subsec:sys_design_sparsity}, we refer to the equalizer as \emph{SS-aware equalizers}, such as SS-MRC and SS-CGA.
Note that LMMSE inherently requires a dense channel matrix $\chMatDDEst$ for the matrix inversion, but SS is designed for efficient MVM {\eqref{eq:sparse_mvm}} only.
For both MRC and CGA, $\chMatDDEst$ is constructed in the pilot frame, and the iterative equalizer is performed for each data frame.
In contrast, LMMSE constructs and inverts $\chMatDDEst$ in the pilot frame and performs one-shot GEMM equalization for each data frame.


\begin{figure}[!t]
    \centering
    \includegraphics[width=0.90\columnwidth]{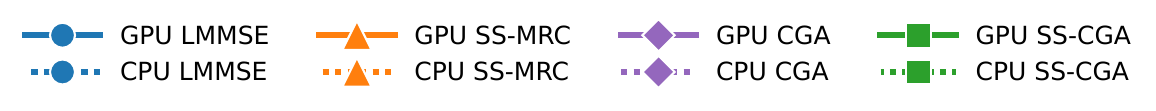}
    \vspace{-3.0mm}
    \includegraphics[width=0.80\columnwidth]{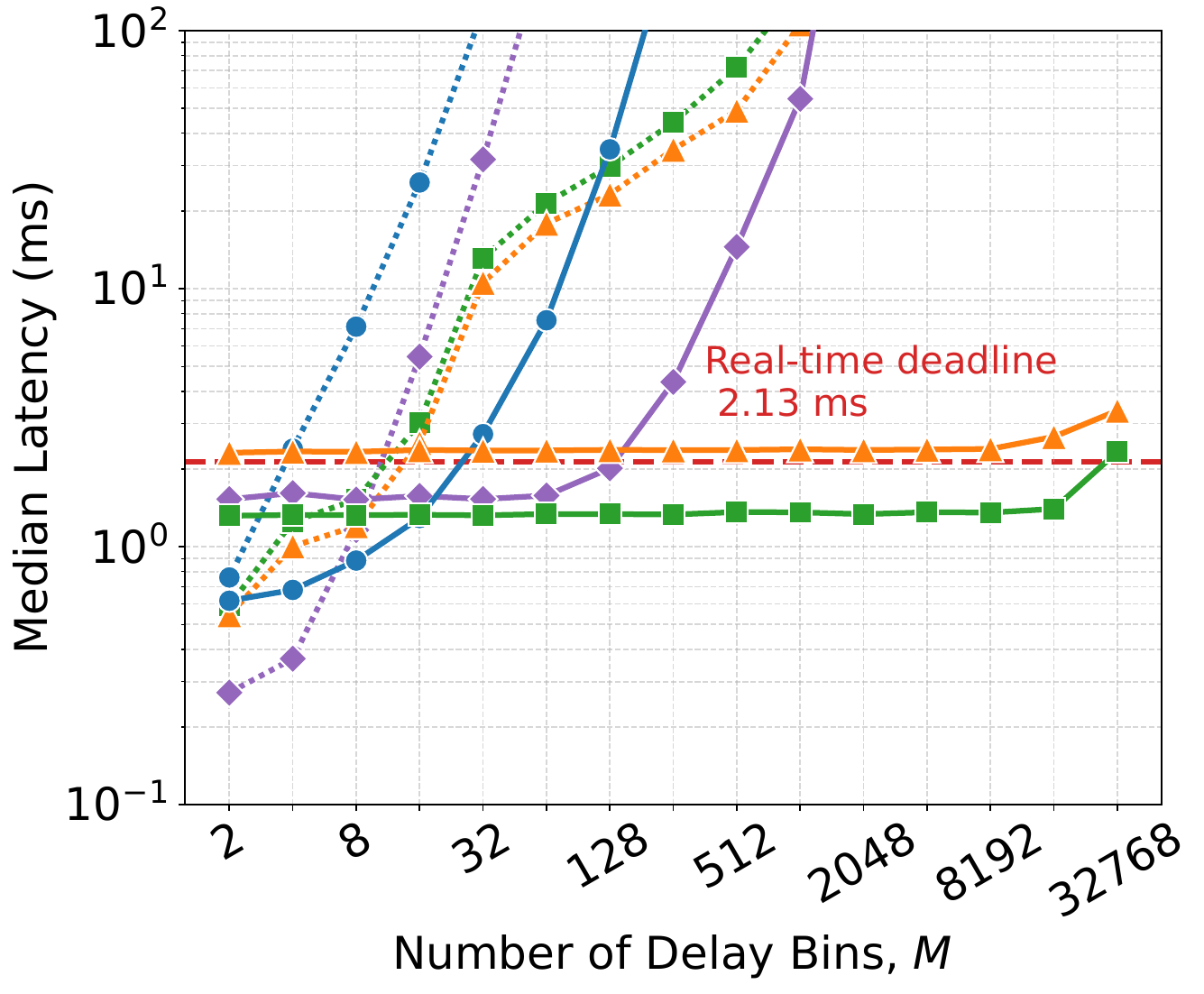}
    \label{subfig:eval_time_vs_m_total}
    \caption{Median end-to-end processing latency across hardware platforms (CPU, GPU), equalizers (LMMSE, MRC, and CGA), and structured-sparsity (SS) awareness, with $\numDopplerBin=32$. 
    The pilot/data frame duration of $\frameTime={1.067}$\thinspace{\msec} and processing deadline is $2\frameTime={2.134}$\thinspace{\msec}.
    }
    \label{fig:eval_time_vs_m_total}
    \vspace{-2.0mm}
\end{figure}

\subsection{End-to-end Performance}
\label{subsec:eval_latency}

A real-time system must process incoming frames within a bounded latency while providing strict timing guarantees.
In this section, we evaluate whether our design satisfies these real-time requirements and quantify the achievable complexity reduction while maintaining BER.


\myparatight{End-to-end processing latency.}
The observed practical scalability in Section~\ref{subsec:scalablity_matop} suggests that MVM demonstrates a constant execution time on GPU within an operation region.
This observation reveals the potential of designing a processing with similar scalability.
Fig.~\ref{fig:eval_time_vs_m_total} shows the end-to-end median latency of the OTFS processing across various compute platforms (CPU and GPU), equalizers (LMMSE, MRC, and CGA), and matrix data structure (with or without leveraging structured sparsity of $\chMatDDEst$).
The end-to-end latency is measured by transmitting an OTFS packet with a point-pilot frame, followed by a data frame, so the packet deadline will be {2.13}\thinspace{\msec}.
For a fair comparison, we let both CGA and MRC perform 10 iterations and a threshold $\heffThres=0.08$ for sparse $\chMatDDEst$ construction in this work, unless otherwise specified.

Fig.~\ref{fig:eval_time_vs_m_total} demonstrates that our system (GPU SS-CGA) can constantly execute the OTFS receiver pipeline within the real-time deadline across the DD domain grid dimension ${\numDelayBin}\times{\numDopplerBin}$.
By leveraging the structured-sparsity data representation, the low-complexity CGA equalizer, and the GPU platform, our system successfully translates the observed unit-test-level scalability of MVM into an end-to-end processing system while meeting the real-time deadline.
The LMMSE-based solutions \eqref{eq:lmmse} suffer from high complexity incurred by GEMM and matrix inversions.
Similarly, the CPU platform is limited by the available hardware parallelism, despite lower latency at smaller grid sizes, leading to suboptimal scalability.
Note that even with low-complexity equalizers (e.g., CGA) and a GPU platform, a data structure exploiting the structured sparsity of $\chMatDDEst$ is the key enabler of constant latency with grid scaling, as it reduces MVM complexity by leveraging the insight that the number of dominant paths is independent of the grid size.


\begin{figure}[!t]
    \centering
    \vspace{-1.5mm}
    \includegraphics[width=0.98\columnwidth]{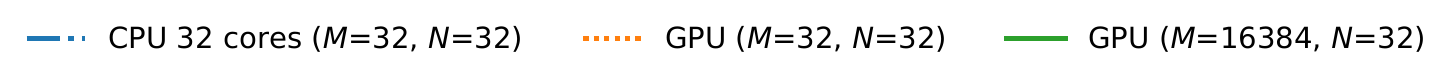}

    \includegraphics[width=0.98\columnwidth]{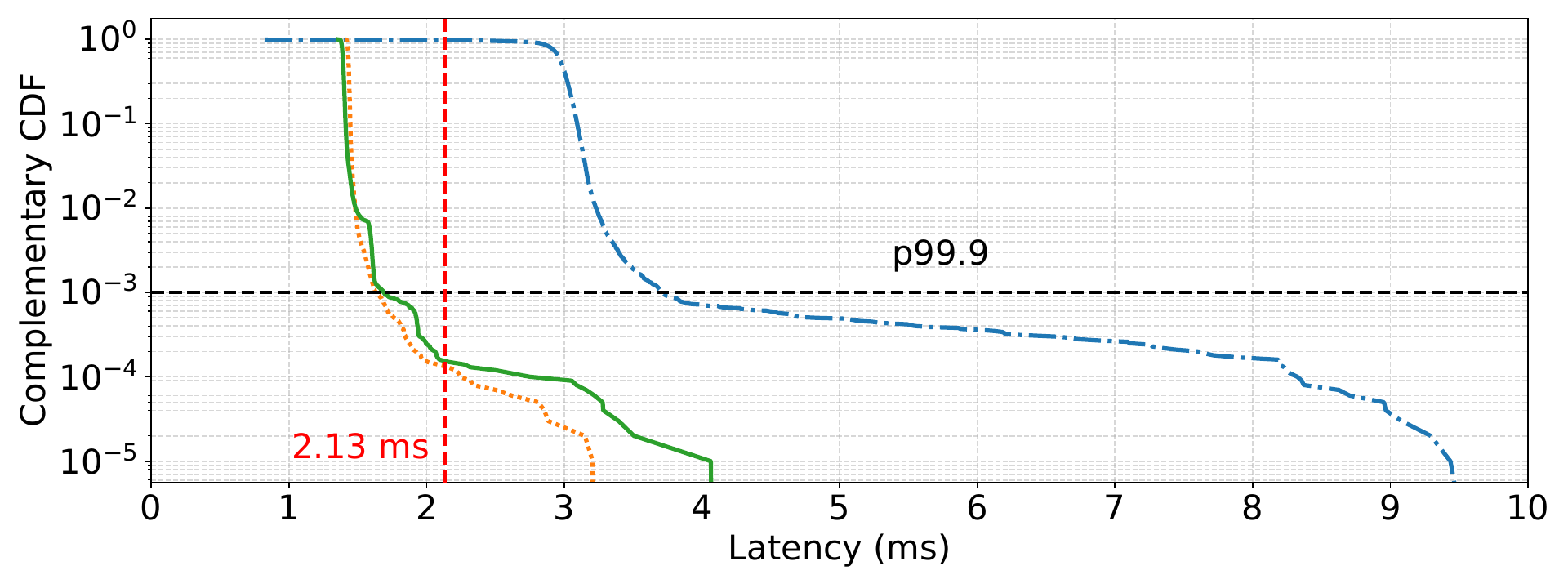}
    \vspace{-3mm}
    \caption{Latency CCDF across 10K packets, where GPU exhibits lower tail latency than the CPU, and increasing the grid size does not significantly affect the distribution.
    Overall, the GPU platform can meet the 99.9-th percentile (p99.9) processing latency for DD grid dimension of up to $(16384,32)$.}
    \label{fig:eval_time_ccdf}
    \vspace{-1.5mm}
\end{figure}

\myparatight{99.9-th percentile processing latency.}
In addition to median processing time, the \emph{deadline satisfaction rate} is a critical metric for evaluating the reliability of a real-time communication link.
Fig.~\ref{fig:eval_time_ccdf} presents the tail latency distribution of the OTFS processing pipeline on both CPU and GPU platforms, evaluated using 10K pilot and data frames.
Due to significantly higher execution time of CPU processing with large grid sizes, we restrict CPU-based evaluation to $({\numDelayBin},{\numDopplerBin}) = (32, 32)$.
On the other hand, GPU-based processing is evaluated under both small and large grid sizes of $({\numDelayBin},{\numDopplerBin}) = (32, 32)$ and $(16384, 32)$.
The results show that GPU processing not only scales efficiently with increasing grid sizes, but also exhibits a tighter latency distribution compared to CPU processing.
Specifically, GPU processing with $({\numDelayBin},{\numDopplerBin}) = (16384, 32)$ can meet the target deadline of {2.13}\thinspace{\msec} at 99.984-th percentile, with a 99.9-th percentile latency of {1.69}\thinspace{\msec}.
In contrast, CPU processing suffers from a larger 99.9-th percentile latency of {3.72}\thinspace{\msec}, approximately 1.75$\times$ higher than the target deadline.


\begin{figure}[!t]
    \centering
    \vspace{-1.5mm}
    \includegraphics[width=0.9\columnwidth]{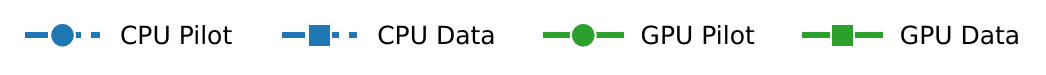}

    \vspace{-4mm}
    \subfloat[LMMSE (Dense)]{
    \includegraphics[width=0.47\columnwidth]{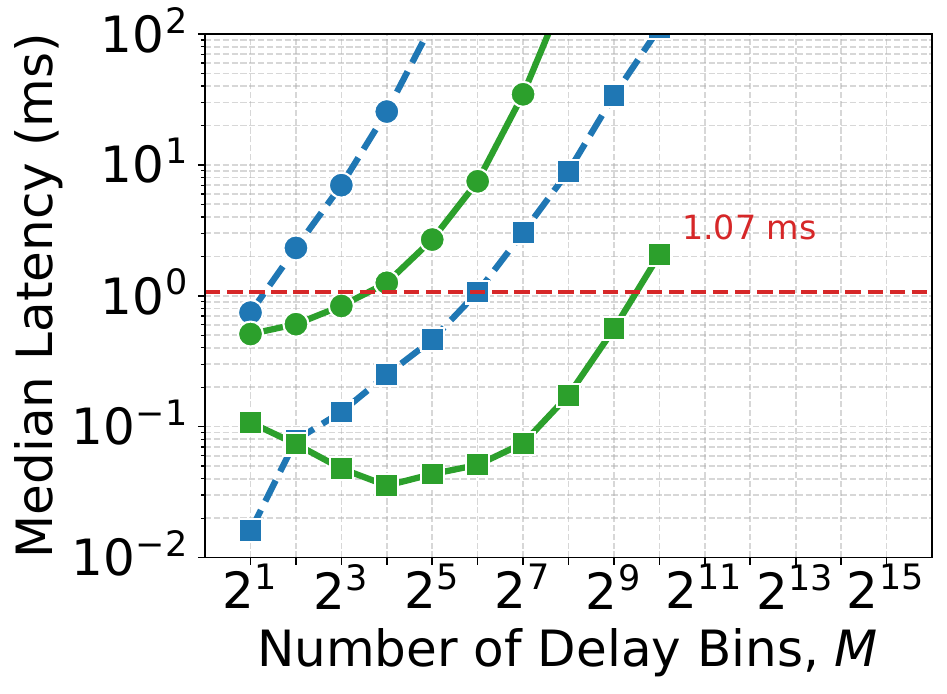}
    \label{subfig:eval_time_vs_m_lmmse_median}}
    \subfloat[SS-MRC]{
    \includegraphics[width=0.47\columnwidth]{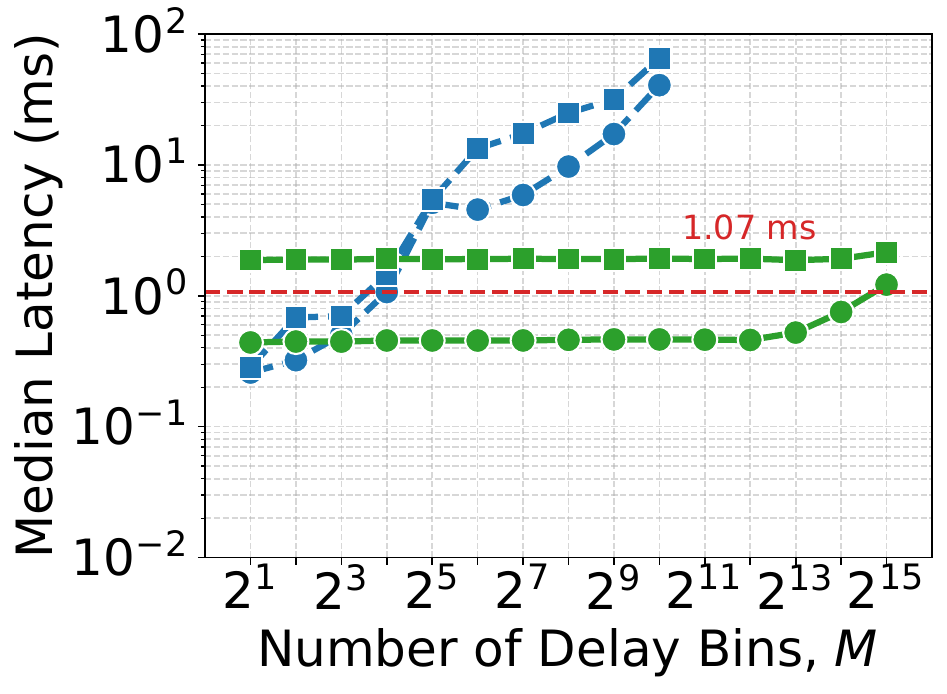}
    \label{subfig:eval_time_vs_m_mrc_median}}

    \vspace{-3mm}
    \subfloat[CGA (Dense)]{
    \includegraphics[width=0.47\columnwidth]{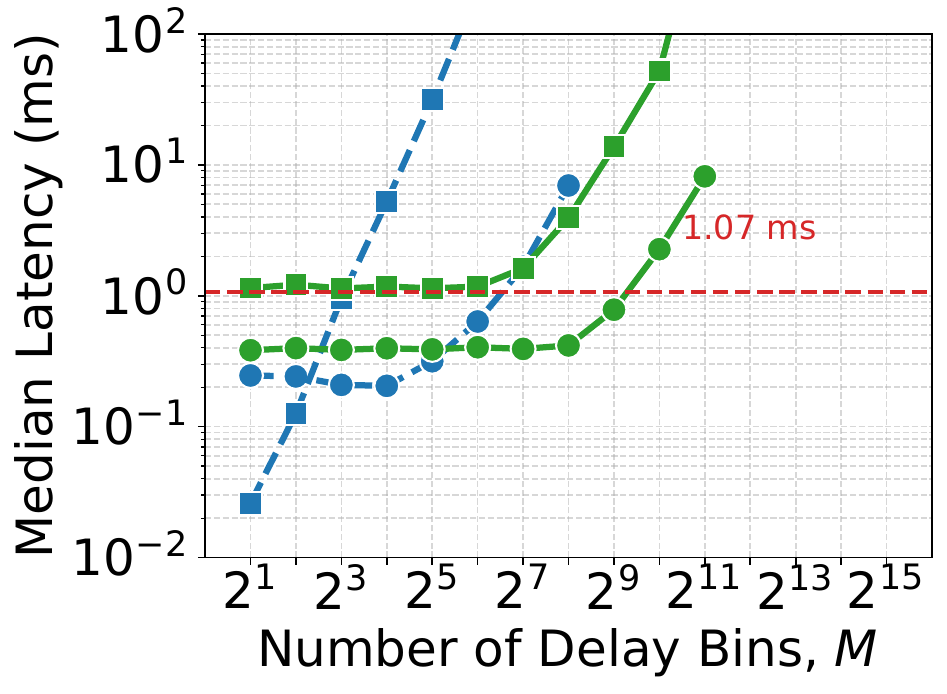}
    \label{subfig:eval_time_vs_m_cga_median}}
    \subfloat[SS-CGA]{
    \includegraphics[width=0.47\columnwidth]{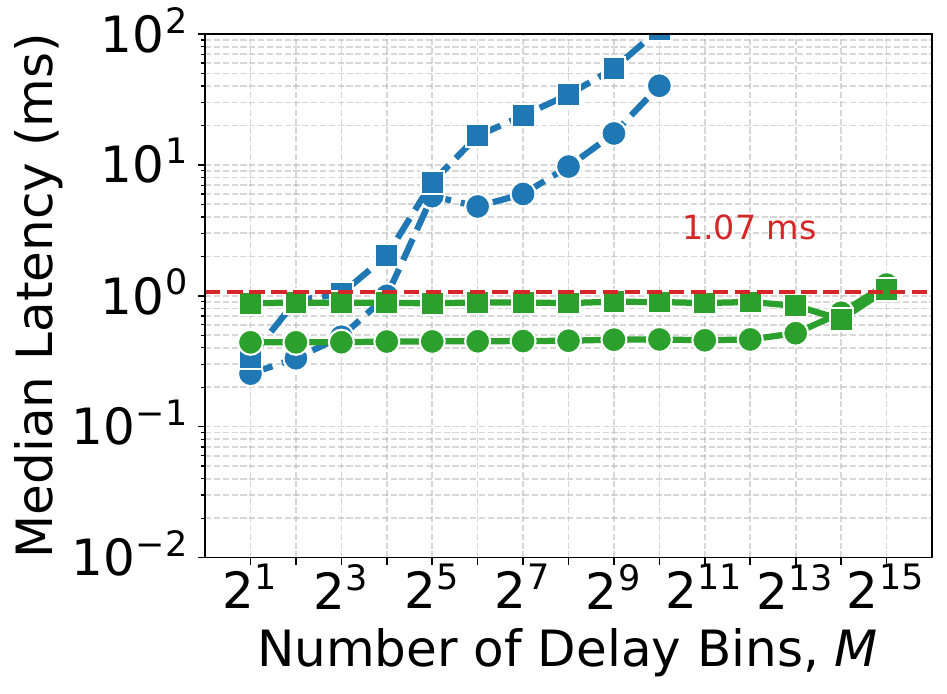}
    \label{subfig:eval_time_vs_m_ss_cga_median}}
    \caption{The pilot and data frame latency breakdown of Fig.~\ref{fig:eval_time_vs_m_total} for each equalizers. Among the two low-complexity equalizers (MRC and CGA), only the proposed SS-CGA constantly meets the latency requirement of {1.07}\thinspace{\msec} for processing both the pilot and the data frame, which is stricter than the packet-level amortized deadline of {2.13}\thinspace{\msec}, on GPU.}
    \label{fig:eval_time_vs_m_eq}
    \vspace{-1.5mm}
\end{figure}

\myparatight{Processing time breakdown.}
Fig.~\ref{fig:eval_time_vs_m_eq} shows the latency of processing pilot and data frames, respectively, for each equalizer in Fig.~\ref{fig:eval_time_vs_m_total} under the same configurations.
The pilot and data frames inherently contain different DSP stages and thus different complexity.
For example, an LMMSE equalizer performs the entire channel estimation in the pilot frame via matrix inversion and requires only one GEMM per data frame, resulting in unbalanced processing time.
In contrast, the MRC and CGA equalizers only construct $\chMatDDEst$ in the pilot frame and iteratively equalize the data frame with more complex algorithms.
Note that even with structured sparsity, GPU SS-MRC exceeds the deadline for data frames, while GPU {\name} meets the deadline for both pilot and data frame constantly for $8\leq\numDelayBin\leq16384$.
%
\begin{figure}[!t]
    \centering
    \vspace{-1mm}
    \includegraphics[width=0.65\columnwidth]{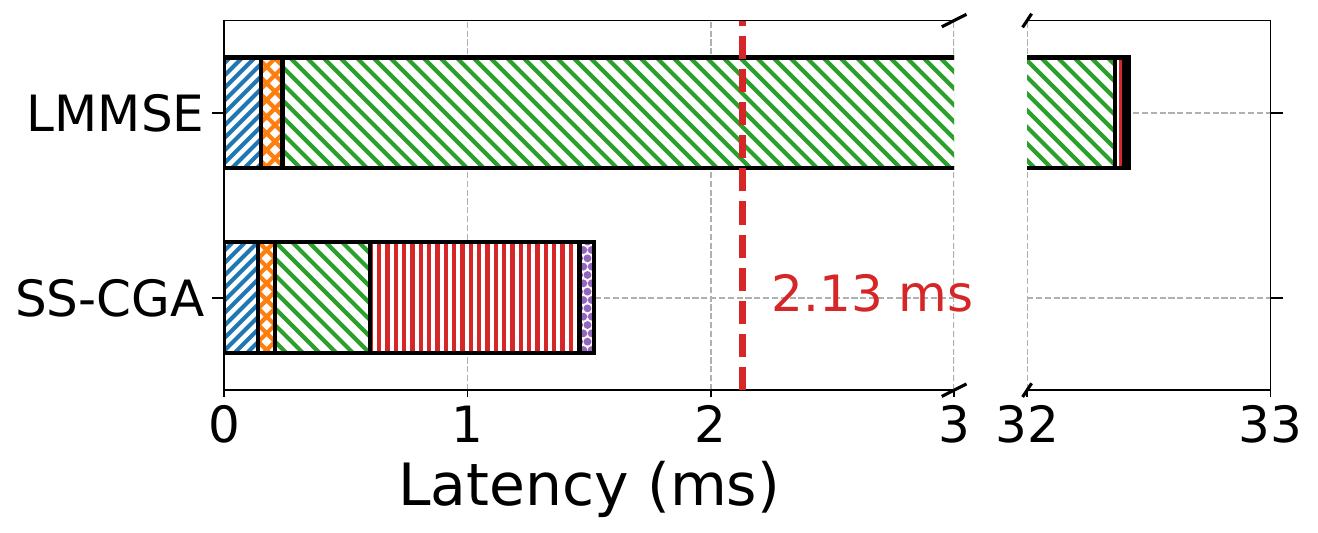}    \hspace{-2mm}
    \subfloat{\raisebox{.38\height}{
        \includegraphics[width=0.22\columnwidth]{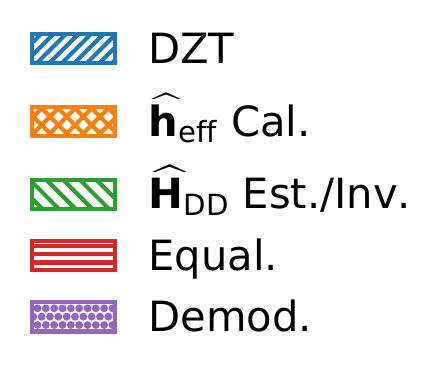}}}
    \vspace{-1.5mm}
    \caption{Breakdown of the 99.9th percentile of the combined one pilot and one data frames with $(\numDelayBin,\numDopplerBin)=(128,32)$ on GPU.
    The LMMSE equalizer suffers from high overhead due to the large matrix inversion in {$\chMatDDEst$}; SS-CGA trades off equalization time while keeping both the pilot and data frames within the per-frame deadline of $\frameTime=1.07$\thinspace{\msec} if viewed separately.}
    \vspace{-1.5mm}
    \label{fig:eval_time_breakdown_combined}
\end{figure}
%
Fig.~\ref{fig:eval_time_breakdown_combined} further compares LMMSE and SS-CGA for the latency of each OTFS processing stage on the GPU in a packet of one pilot and one data frame.
SS-CGA perform Algo.~\ref{alg:dia_sparse_cga} with MVM in each iteration, leading to {0.86}\thinspace{\msec} equalization time compared to the one-shot GEMM of LMMSE in {0.04}\thinspace{\msec}.
However, the matrix inversion in LMMSE incurs significant overhead in pilot-frame processing (i.e., {31.87}\thinspace{\msec} for $\chMatDDEst$ inversion), preventing practical deployment.
In contrast, the SS-CGA constructs \eqref{eq:ds_repr} in {0.34}\thinspace{\msec}.
Note that, despite the longer equalization, SS-CGA meets the frame deadline for both the pilot and data frames, demonstrating a preferable design trade-off, and maintains a good scalability due to structured sparsity, as shown in Fig.~\ref{fig:eval_time_vs_m_total}.


\begin{figure}[!t]
    \centering
    \vspace{-2.0mm}
    \includegraphics[width=0.6\columnwidth]{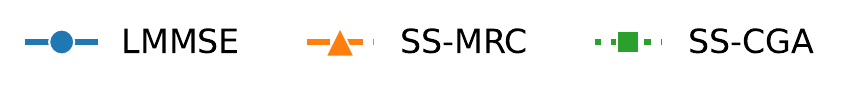}\vspace{-4mm}
    \subfloat[BER vs. SNR (QPSK)]{
    \includegraphics[width=0.45\columnwidth]{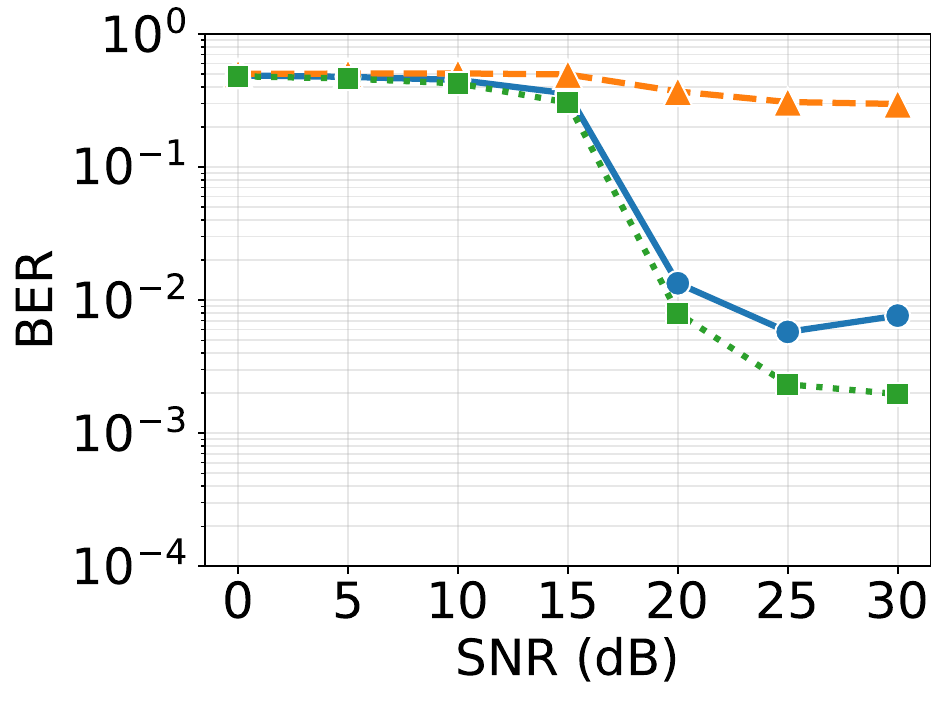}
    \label{subfig:eval_ber_vs_snr_qpsk}}
    \subfloat[BER vs. $\chMaxDopplerFreq$ (QPSK)]{
    \includegraphics[width=0.45\columnwidth]{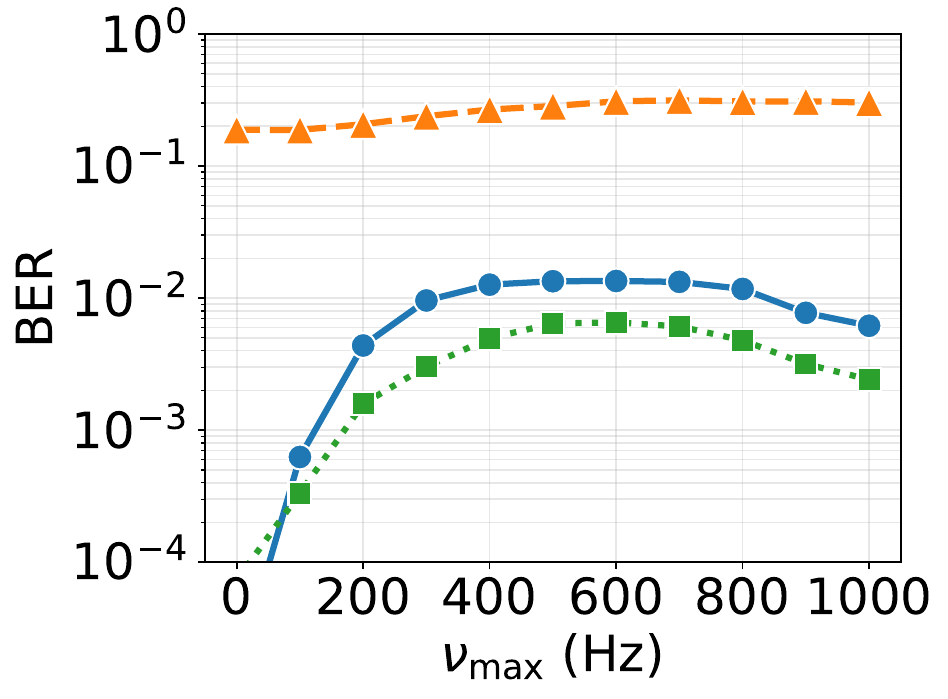}
    \label{subfig:eval_ber_vs_doppler_qpsk}}
    \vspace{-4mm}
    \subfloat[BER vs. SNR (16QAM)]{
    \includegraphics[width=0.45\columnwidth]{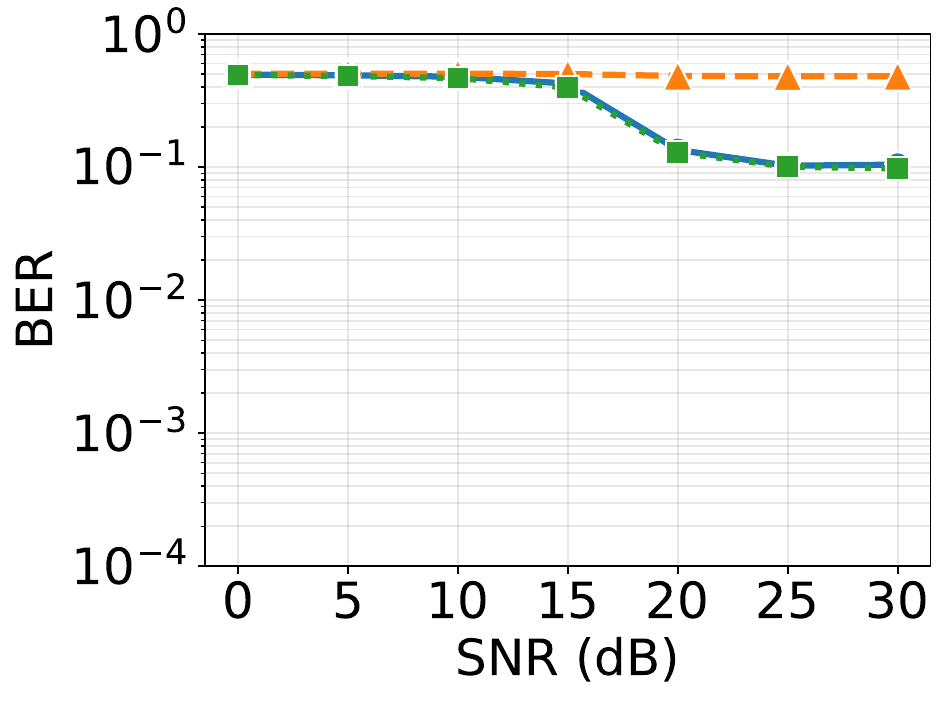}
    \label{subfig:eval_ber_vs_snr_qam16}}
    \subfloat[BER vs. $\chMaxDopplerFreq$ (16QAM)]{
    \includegraphics[width=0.45\columnwidth]{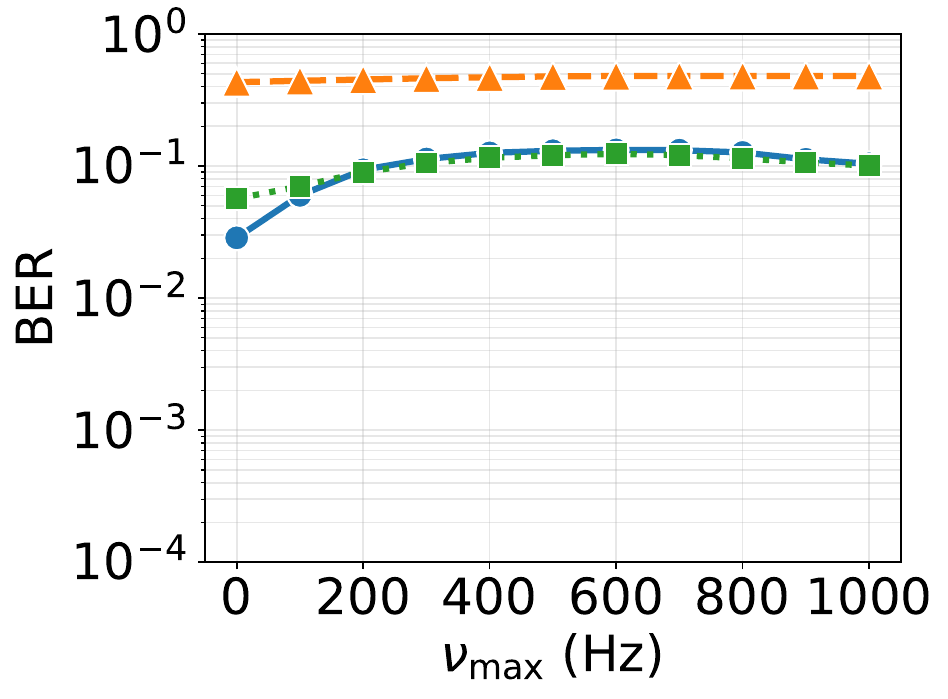}
    \label{subfig:eval_ber_vs_doppler_qam16}}
    \caption{
    BER vs. SNR at $\chMaxDopplerFreq=100$\thinspace{Hz} and BER vs. $\chMaxDopplerFreq$ at {25}\thinspace{dB} SNR across equalizers and QAM scheme. 
    ($\numDelayBin$, $\numDopplerBin$) = (128, 32), and $\heffThres$=0.08.
    SS-CGA outperforms MRC and matches LMMSE in BER.
    }
    \label{fig:eval_ber_vs_ch}
    \vspace{-1.5mm}
\end{figure}

\begin{figure}[!t]
    \centering
    \vspace{-1.5mm}
    \subfloat[BER vs. $\numDelayBin$ at $\chMaxDopplerFreq=100$\thinspace{Hz}]{
    \includegraphics[width=0.45\columnwidth]{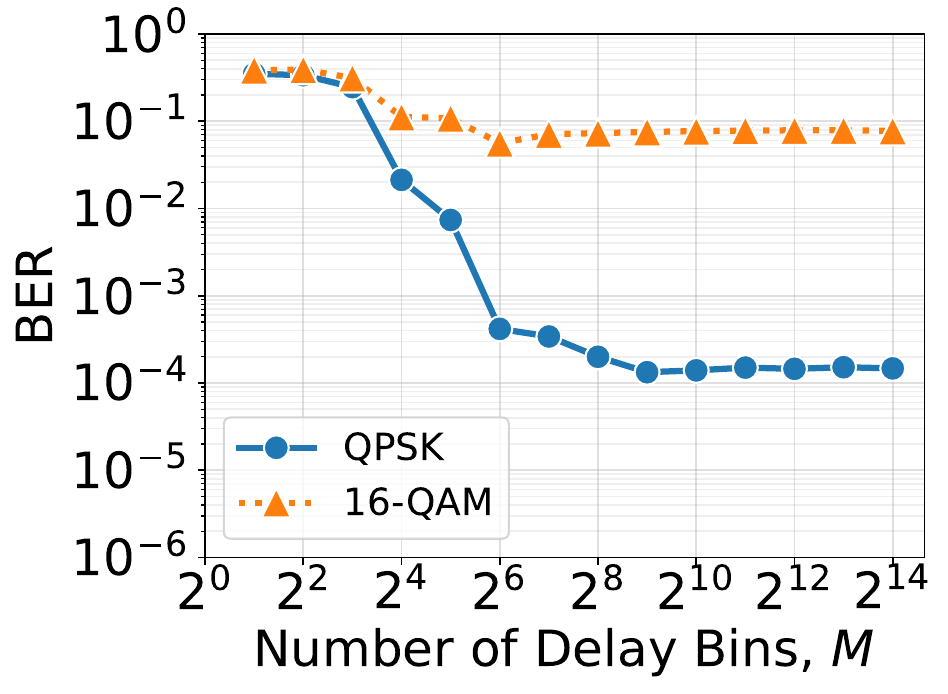}
    \label{subfig:ber_vs_M}}
    \subfloat[BER vs. $\chMaxDopplerFreq$ (QPSK)]{
    \includegraphics[width=0.45\columnwidth]{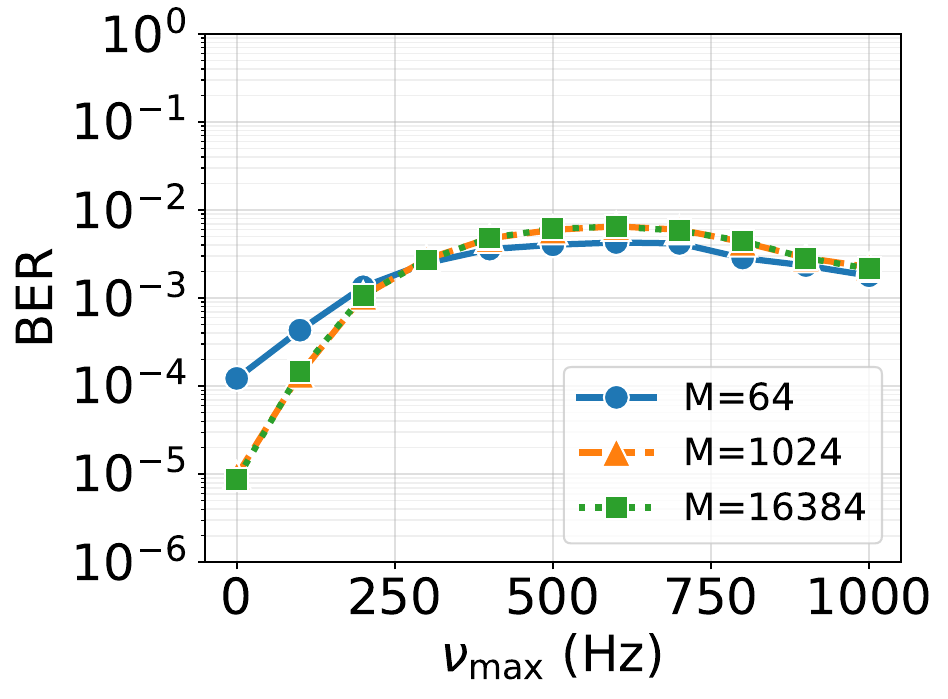}
    \label{subfig:ber_vs_numax_by_M}}
    \caption{
    BER of SS-CGA with {25}\thinspace{dB} SNR.
    BER decreases when $\numDelayBin$ increases, with QPSK having lower BER.
    Across multiple $\numDelayBin$ values, the BER increases with $\chMaxDopplerFreq$ but SS-CGA still maintains BER lower than 1\%.
    }
    \label{fig:ber_vs_M_numax}
    \vspace{-1.5mm}
\end{figure}

\myparatight{BER.}
We also evaluate BER versus SNR and ${\chMaxDopplerFreq}$ across combinations of equalizer design (LMMSE, SS-MRC, and SS-CGA) and QAM modulation (QPSK and 16QAM) using the Veh-A channel (Section~\ref{sec:impl}).
SNR is swept from 0 to {30}\thinspace{dB}, and each operating point is averaged over 1,000 randomly generated bitstreams and independent Veh-A channel realizations with AWGN.
For the Veh-A channel model, the maximum Doppler spread $\chMaxDopplerFreq$ ranges from 0 to {1000}\thinspace{Hz} in {100}\thinspace{Hz} increments.
Fig.~\ref{fig:eval_ber_vs_ch} shows that the proposed SS-CGA achieves similar or better BER than LMMSE, while constantly outperforms SS-MRC.
These results verify that the proposed structured-sparsity representation using dominant paths with an iterative low-complexity equalizer does not sacrifice communication quality.
Overall, BER decreases in all cases when SNR increases or ${\chMaxDopplerFreq}$ decreases, as higher noise power and higher Doppler shift degrade the detection capability of the receiver pipeline.
As expected, 16QAM exhibits higher BER across all equalizers due to its greater sensitivity to noise.

Fig.~\ref{fig:ber_vs_M_numax} reports the BER for SS-CGA with larger OTFS grid sizes by varying the number of delay bins $\numDelayBin$ under {25}\thinspace{dB} SNR.
Fig.~\ref{fig:ber_vs_M_numax}\subref{subfig:ber_vs_M} shows BER over $\numDelayBin$ under $\chMaxDopplerFreq=100$\thinspace{Hz} for QPSK and 16QAM.
BER first decreases when $\numDelayBin$ increases since a larger $\numDelayBin$ provides better delay resolution $\delayRes$.
After all the paths are separable around $\numDelayBin=64$, increasing $\numDelayBin$ starts to have diminishing gain on BER; BER saturates around $10^{-4}$ for QPSK.
The 16QAM trend aligns with QPSK but with a higher BER due to its greater sensitivity to noise.
Fig.~\ref{fig:ber_vs_M_numax}\subref{subfig:ber_vs_numax_by_M} shows BER over a range of Doppler spread $\chMaxDopplerFreq$ from {0--1000}\thinspace{Hz} for QPSK across three $\numDelayBin$ values.
As expected, BER degrades with larger $\chMaxDopplerFreq$ since the number of Doppler bins $\numDopplerBin$ does not change in the experiment.
However, SS-CGA still maintains BER below 1\% across the three cases.

\subsection{BER, Sparsity, and Processing Latency Trade-offs}


\begin{figure}[!t]
    \centering
    \vspace{-1.5mm}
    \subfloat{
    \includegraphics[width=0.80\columnwidth]{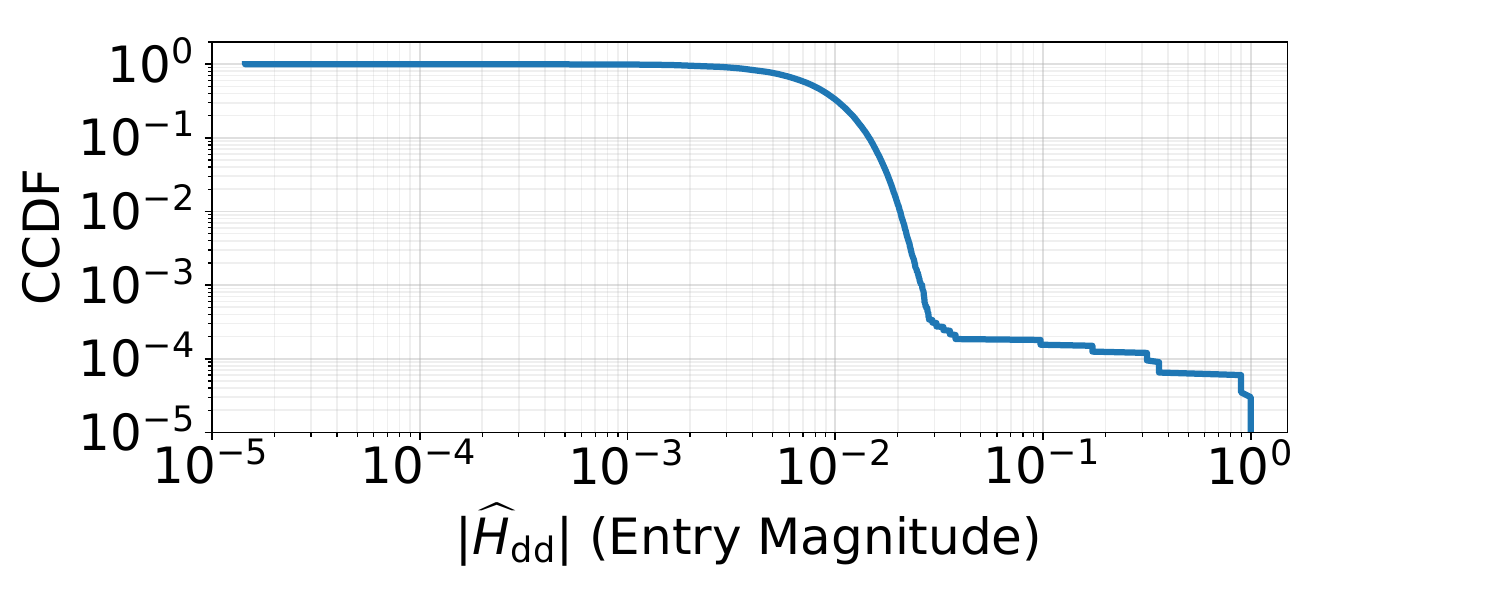}
    \label{subfig:eval_threshold_sweep_ccdf_fast}}
    \vspace{0mm}
    \subfloat{
    \includegraphics[width=0.80\columnwidth]{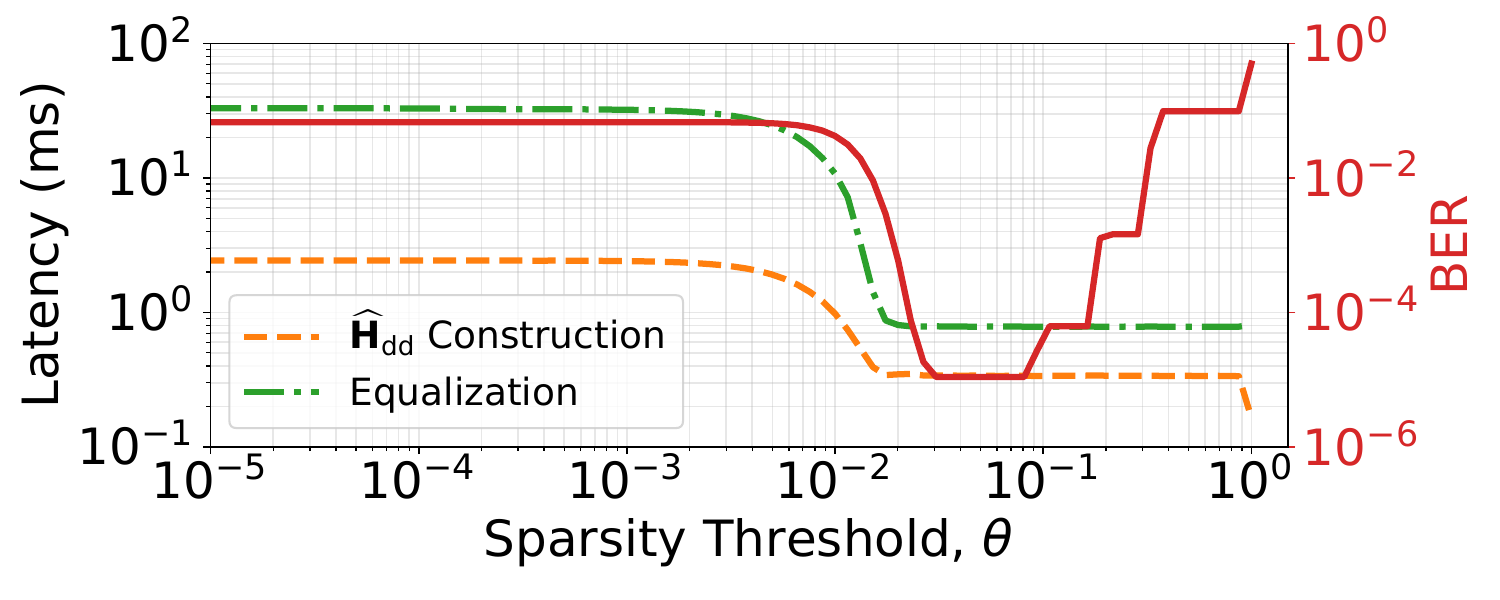}
    \label{subfig:eval_threshold_sweep_time_ber_combined}}
    \vspace{-3.0mm}
    \caption{The magnitude distribution of $\chMatDDEst$ elements (top) and the execution time/BER vs. sparsity thresholds $\heffThres$ (bottom) with $(\numDelayBin,\numDopplerBin)=(128,32)$, QPSK, SNR = {30}\thinspace{dB}.
    An operating region is observed for $0.03 \leq \heffThres \leq 0.08$.}
    \label{fig:eval_threshold_sweep}
    \vspace{-2.0mm}
\end{figure}

In Section~\ref{subsec:sys_design_sparsity}, we discuss how our system leverages the sparsity of $\chMatDDEst$ from the $\numPath$ dominant paths, which can be obtained via a threshold $\heffThres$ on $\chEffEstMat$ \eqref{eq:heff_thres}.
We hereby provide an example-based numerical analysis for the three-way trade-off among the $\chMatDDEst$ sparsity, the processing latency, and the final BER, as functions of $\heffThres$.
Fig.~\ref{fig:eval_threshold_sweep} reports
\emph{(i)} the absolute magnitude CCDF of elements in $\chMatDDEst$,
\emph{(ii)} the average (100 trials) processing latency of both $\chMatDDEst$ construction and equalization of our system on GPU, and
\emph{(iii)} the final resultant BER,
by sweeping the threshold $\heffThres$.
The magnitude of elements in $\chMatDDEst$ is normalized to one, and two clusters around $0.02$ and $1$ are presented.
Hence, the thresholds between that range (i.e., $0.03 \leq \heffThres \leq 0.08$) lead to a desired tradeoff.
For example, at $\heffThres = 0.08$, compared to the unthresholded case,
\emph{(i)} 99.98\% of the $\chMatDDEst$ elements can be pruned, so that
\emph{(ii)} the $\chMatDDEst$ construction speeds up from {2.43}\thinspace{\msec} to {0.34}\thinspace{\msec} (7.18$\times$) and equalization from {32.92}\thinspace{\msec} to {0.78}\thinspace{\msec} (297.31$\times$), while
\emph{(iii)} the BER is significantly improved from 0.067\% to 0.001\%.
This example suggests that a proper threshold $\heffThres$ benefits both the system processing latency and the BER.
This threshold $\heffThres$ can be empirically selected based on the SNR estimated from the pilot frame, inferred from the number of reflectors in the channel, if known, or based on the desired latency requirement.

\subsection{BER Convergence in CGA Equalizer}\label{subsec:cga_ber_converge}

\begin{figure}[!t]
    \centering
    \vspace{-1.5mm}
    \includegraphics[width=0.9\columnwidth]{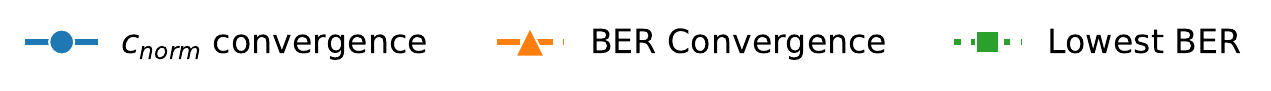}\vspace{-4mm}
    \subfloat[$\numDelayBin=128$, $\numDopplerBin=32$]{
    \includegraphics[width=0.45\columnwidth]{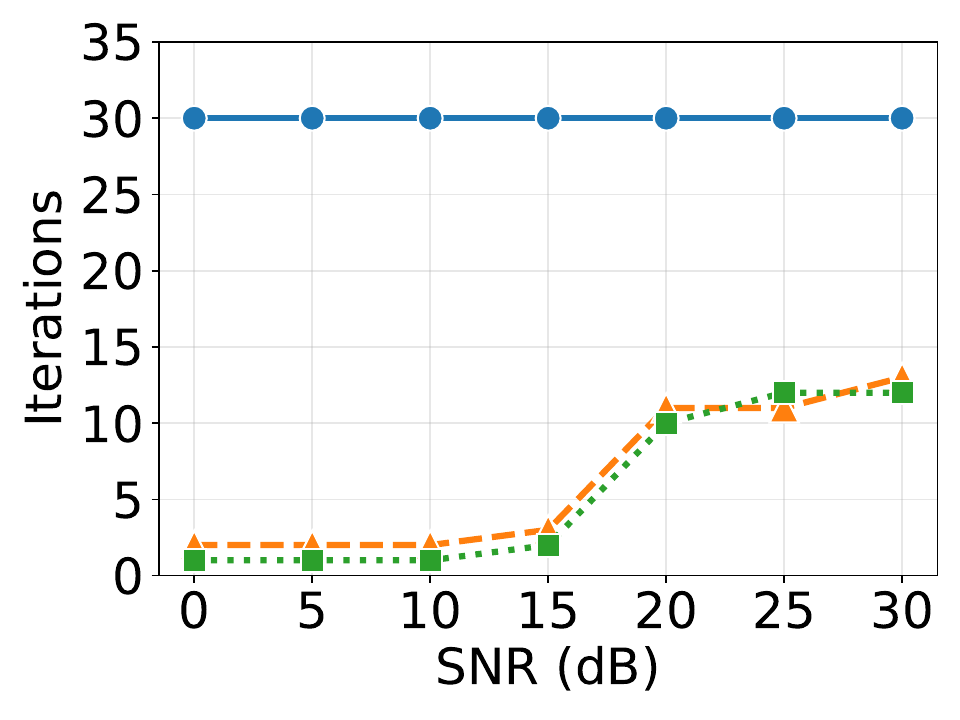}
    \label{subfig:eval_cga_iters_vs_snr_M128_N32}}
    \subfloat[$\numDelayBin=\numDopplerBin=32$]{
    \includegraphics[width=0.45\columnwidth]{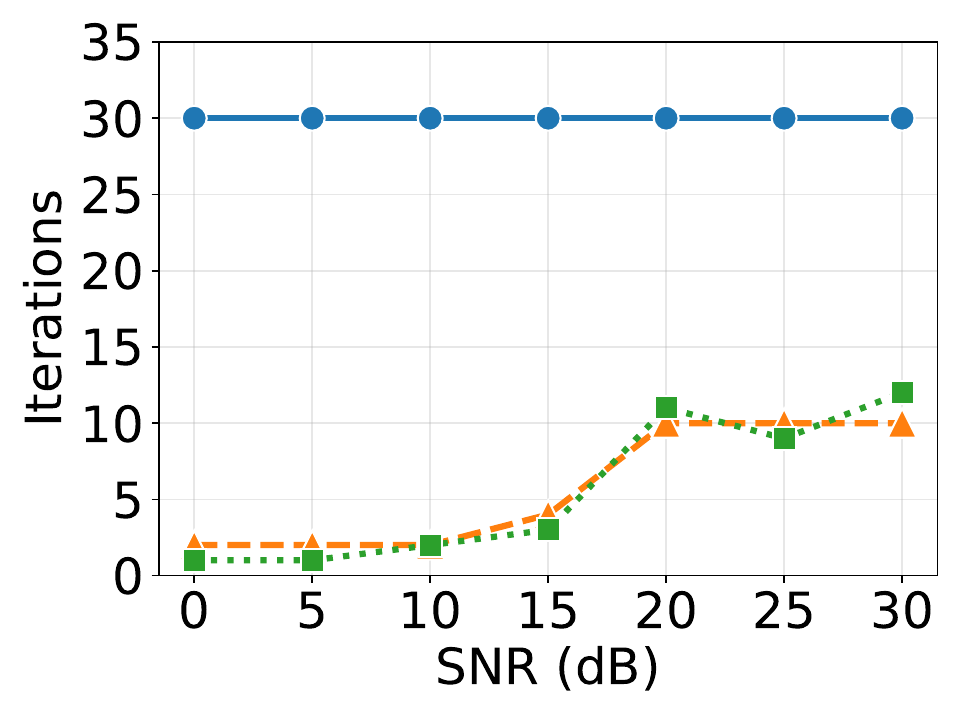}
    \label{subfig:eval_cga_iters_vs_snr_M32_N32}}
    \caption{The iterations in Algo.~\ref{alg:dia_sparse_cga} across SNR at the original $\sqResNorm$-based termination~\cite{mattu2025low}, the lowest BER, and the BER convergence \eqref{eq:ber_convergence}.
    The lowest BER and BER convergence lines are generally close.
    In contrast, $\sqResNorm$-based termination constantly exceeds the designed maximum of 30 iterations and requires many more iterations after the lowest BER is achieved.
    }
    \label{fig:eval_cga_iter_vs_snr}
    \vspace{-1.5mm}
\end{figure}

\begin{figure}[!t]
    \centering
    \vspace{-1.5mm}
    \includegraphics[width=0.9\columnwidth]{figs/eval_cga_iters_vs_snr_legend.pdf}\vspace{-4mm}
    \subfloat[$\numDelayBin=128$, $\numDopplerBin=32$]{
    \includegraphics[width=0.45\columnwidth]{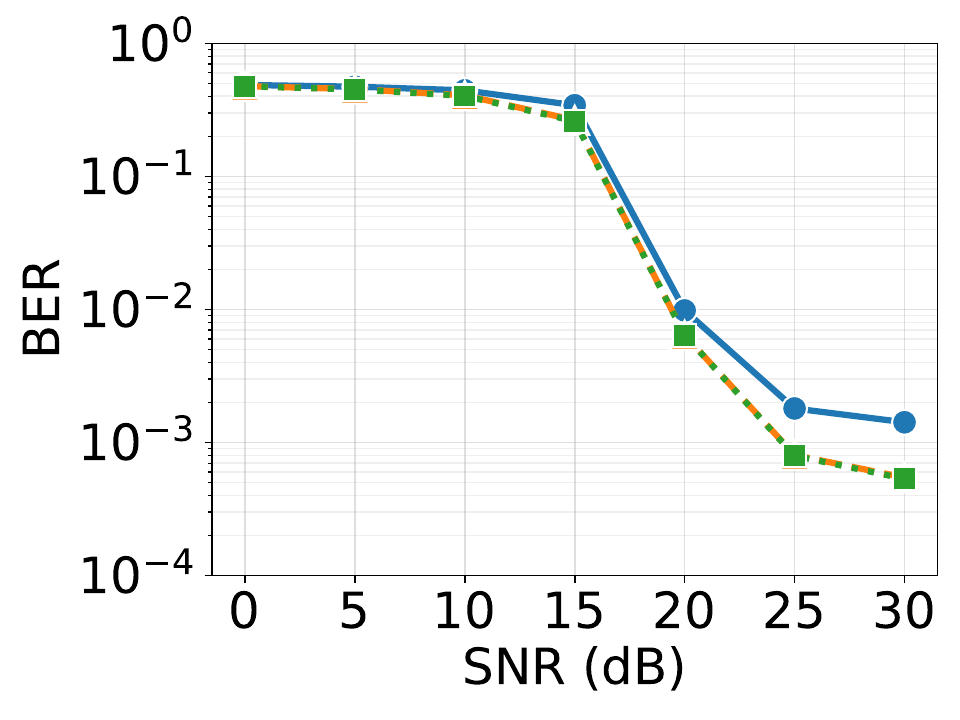}
    \label{subfig:eval_cga_ber_vs_snr_M128_N32}}
    \subfloat[$\numDelayBin=\numDopplerBin=32$]{
    \includegraphics[width=0.45\columnwidth]{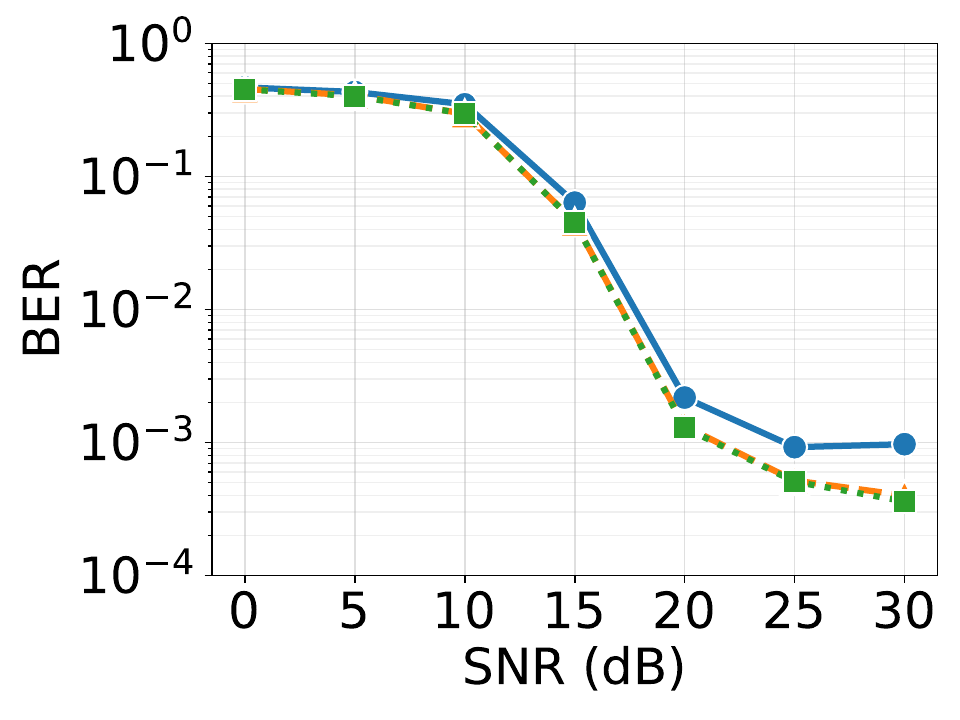}
    \label{subfig:eval_cga_ber_vs_snr_M32_N32}}
    \caption{The achieved BER at the iteration of the original $\sqResNorm$-based termination~\cite{mattu2025low}, the measured lowest BER, and the converged BER \eqref{eq:ber_convergence} in Algo.~\ref{alg:dia_sparse_cga} across SNR.
    $\sqResNorm$-based termination constantly yields higher BER.}
    \label{fig:eval_cga_ber_vs_snr}
    \vspace{-2.0mm}
\end{figure}

The unconditional termination (Section~\ref{subsec:sys_design_cga}) requires offline profiled iteration counts to guarantee the desired algorithm convergence.
In Fig.~\ref{fig:eval_cga_ber_cnorm_vs_iter_M128_N32}, the trend of $\sqResNorm$ decreasing over iteration is shown.
This section compares the behavior and BER of $\sqResNorm$-monitoring and the offline-profiling methods using the unconditional CGA (Algo.~\ref{alg:dia_sparse_cga}).
In each experiment, we run Algo.~\ref{alg:dia_sparse_cga} up to 30 iterations and record the iterations and BER at which
\emph{(i)} the $\sqResNorm$ converges based on a fixed threshold $\epsilon^2$~\cite{mattu2025low},
\emph{(ii)} the lowest BER is achieved, and
\emph{(iii)} the BER convergence iteration defined in Section~\ref{subsec:sys_design_cga}.

\myparatight{Iterations of BER and CGA convergence.}
Fig.~\ref{fig:eval_cga_iter_vs_snr} shows the recorded iterations for $\sqResNorm$ convergence, BER convergence, and the lowest (best) BER on $(\numDelayBin, \numDopplerBin) = (128, 32)$ and $(32, 32)$, varying SNR.
The results reveal that $\sqResNorm$ cannot fall below its $\epsilon^2=10^{-3}$ threshold~\cite{mattu2025low} and reaches the maximum iterations in all cases.
This is because $\sqResNorm$ scales linearly with $\numDelayBin\numDopplerBin$ in Algo.~\ref{alg:dia_sparse_cga} (line~\ref{line:c_norm_compute}) and experiences diminishing slope despite monotonic decay over iterations (Fig.~\ref{fig:eval_cga_ber_cnorm_vs_iter_M128_N32}).
In contrast, both the BER convergence \eqref{eq:ber_convergence} and the lowest BER appear in much earlier iterations, indicating that the additional iterations yield diminishing BER gain if the equalizer uses $\sqResNorm$ as the termination indicator.
Instead, our offline profiling empirically suggests that CGA converges fast in early iterations.

\myparatight{BER.}
Fig.~\ref{fig:eval_cga_ber_vs_snr} shows the BER corresponding to the recorded iterations in Fig.~\ref{fig:eval_cga_iter_vs_snr} and that achieving the converged BER and the lowest BER requires consistently lower iterations than the maximum iterations (which is also where the $\sqResNorm$-based termination), indicating that additional iterations after BER convergence are unnecessary without BER improvement.
This observation demonstrates the effectiveness of the proposed BER convergence criterion {\eqref{eq:ber_convergence}} since the converged BER overlaps with the lowest BER.
Hence, we set the iteration count $\cgaIters=10$ in Algo.~\ref{alg:dia_sparse_cga} for all SNRs in our evaluation.

\subsection{Latency Requirement and GPU Platform Capability}

\begin{figure}[!t]
    \centering
    \vspace{-1.5mm}
    \subfloat[$\numDelayBin$=1024 across GPUs]{
    \includegraphics[width=0.45\columnwidth]{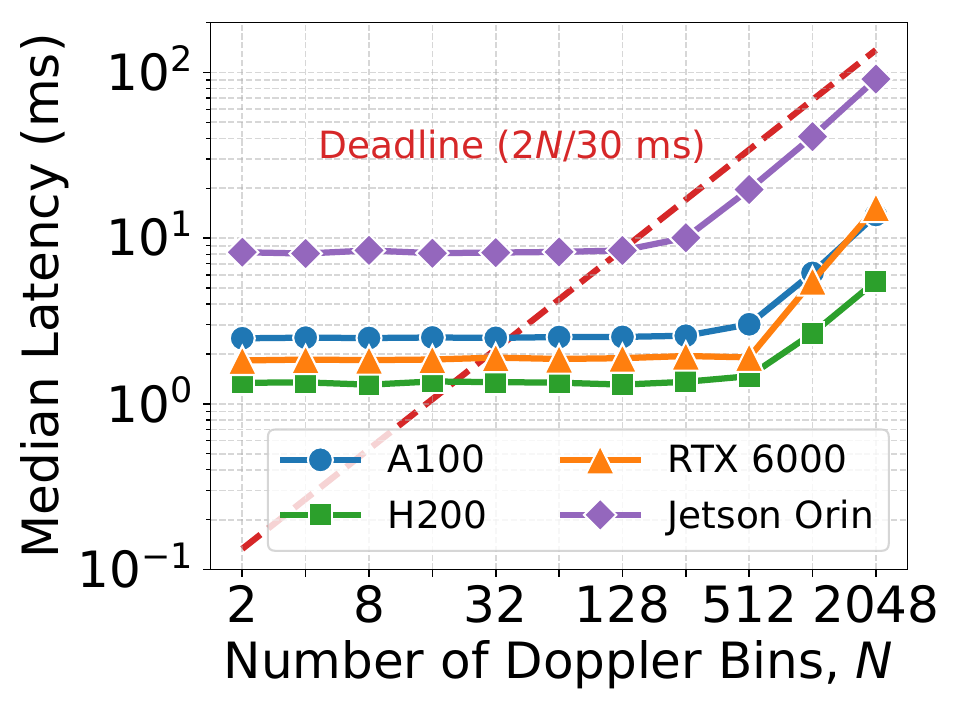}
    \label{subfig:eval_e2e_time_vs_N_cross_platform_median}}
    \subfloat[Jetson Orin across $\numDelayBin$]{
    \includegraphics[width=0.45\columnwidth]{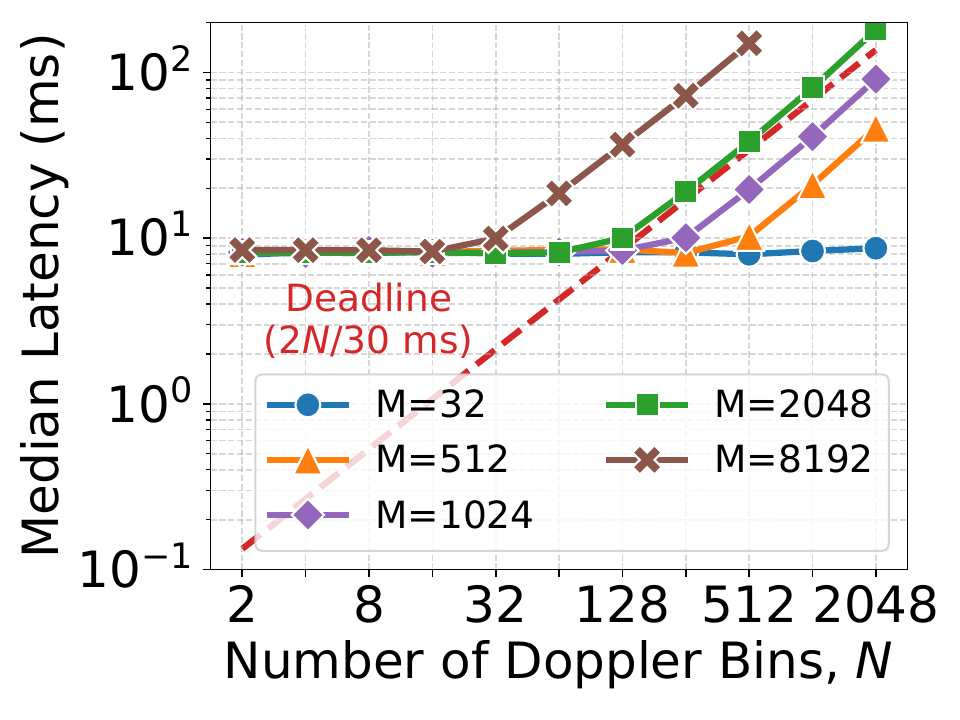}
    \label{subfig:eval_e2e_time_vs_N_jetson_by_M_median}}
    \caption{Median processing latency across GPUs and DD grid sizes $(\numDelayBin,\numDopplerBin)$, with (a) varying $\numDopplerBin$ (which affects the processing deadline), and
    (b) varying both $\numDelayBin$ and $\numDopplerBin$ on the Jetson Orin platform.}
    \label{fig:eval_e2e_time_vs_N_cross_platform}
    \vspace{-1.5mm}
\end{figure}

\begin{figure}[!t]
    \centering
    \vspace{-1.5mm}
    \includegraphics[width=0.8\columnwidth]{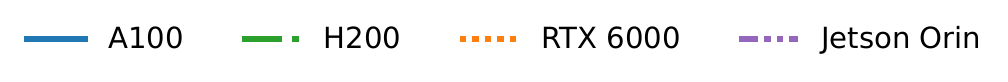}
    \vspace{-1.5mm}

    \includegraphics[width=0.9\columnwidth]{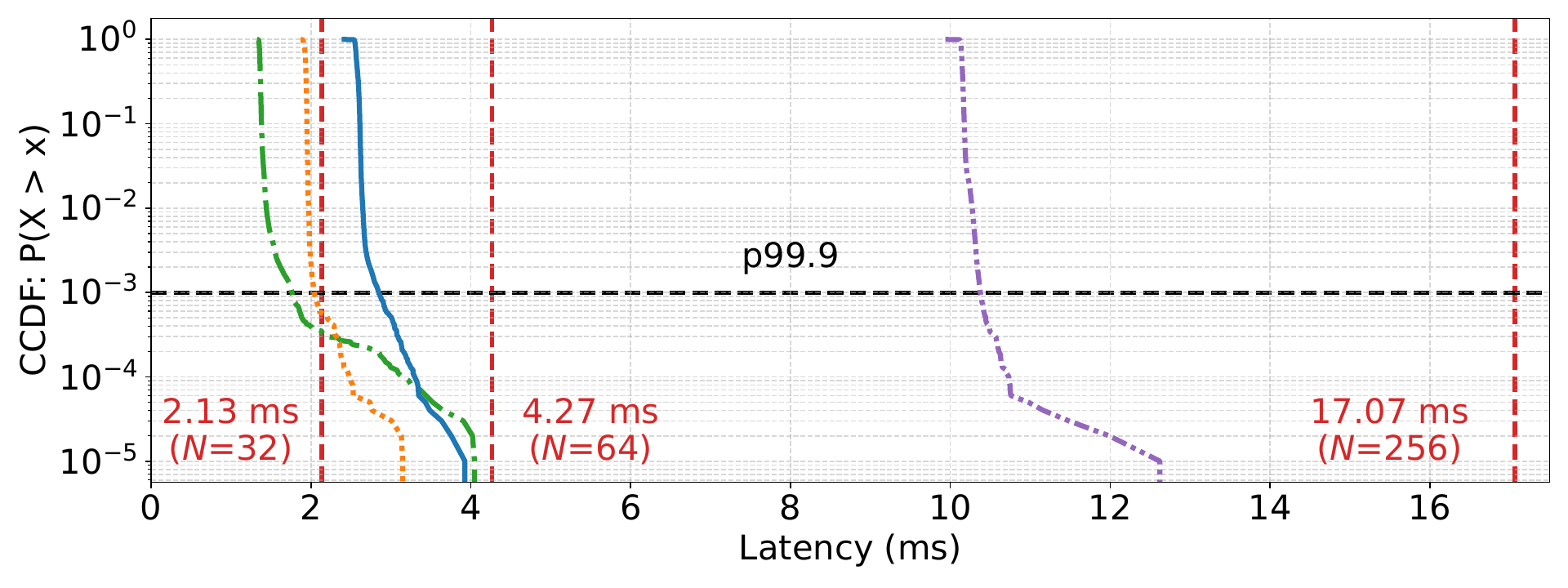}
    \vspace{-3mm}
    \caption{Complementary CDF of the processing latency CCDF across four GPU platforms.
    Each GPU is evaluated with $\numDelayBin=1024$ and the largest $\numDopplerBin$ value that satisfies the 99.9\% latency deadline: $\numDopplerBin = 32$ for RTX~6000 and H200, $\numDopplerBin = 64$ for A100, and $\numDopplerBin = 256$ for Jetson Orin.
    }
    \label{fig:eval_time_ccdf_plat}
    \vspace{-1.5mm}
\end{figure}

The evaluations so far assumed a fixed $\numDopplerBin$ as it corresponds to the frame duration $\frameTime$ and thus the processing deadline {\eqref{eq:bandwidth_frametime}}.
However, GPU platforms exhibit slightly different latencies in unit tests of matrix operations in Section~\ref{subsec:scalablity_matop}.
In this section, we evaluate how different GPU platforms affect the processing latency under varying values of $\numDopplerBin$.
Varying $\numDopplerBin$ does not change the system throughput when $\numDelayBin$ and $\freqSpacing$ remain constant~\eqref{eq:bandwidth_frametime}.

Fig.~\ref{fig:eval_e2e_time_vs_N_cross_platform}\subref{subfig:eval_e2e_time_vs_N_cross_platform_median} shows the median end-to-end latency across 10K packets across four GPU platforms (NVIDIA Jetson Orin, RTX 6000 Ada, A100, H200), with $\numDelayBin=1024$ and varying $\numDopplerBin$.
Given $\freqSpacing=30$\thinspace{kHz}, each pilot/data frame has a duration of $2\frameTime$, with
$\frameTime=\tfrac{\numDopplerBin}{\freqSpacing} = \tfrac{\numDopplerBin}{30\thinspace{\text{kHz}}} (\text{s}) = \tfrac{\numDopplerBin}{30\thinspace{\text{Hz}}} (\text{\msec})$.
GPU platforms have individual constant overhead and upward turning points, resulting in intersections of latency trends to the real-time processing deadline $2\frameTime$.
These intersections mark their ability to satisfy the tighter deadline constraints with a smaller $\numDopplerBin$.
The upward turning points mark where the size of the OTFS grid $(\numDelayBin,\numDopplerBin)$ starts to exceed the parallelization capacity of a GPU and leads to latency scaling.
Hence, Jetson Orin, which is designed for edge computing with lower hardware specifications, exhibits higher constant overhead and an earlier upward turning point.
Note that as $\numDelayBin$ increases, the latency lines may not have intersections with the processing deadline line, implying deadline violation for all possible $\numDopplerBin$ values.

Fig.~\ref{fig:eval_e2e_time_vs_N_cross_platform}\subref{subfig:eval_e2e_time_vs_N_jetson_by_M_median} shows the median latency across 10K packets of Jetson Orin for multiple $\numDelayBin$ values with varying $\numDopplerBin$.
As $\numDelayBin$ increases, the upward turning points move leftward, as product $\numDelayBin\times\numDopplerBin$ remains constant.
Hence, when $\numDelayBin$ is sufficiently large ($\numDelayBin\geq2048$ for Jetson Orin), a GPU will never be able to meet the real-time deadline, marking its capability limit.
Recall that a higher $\numDelayBin$ corresponds to a higher input sample rate when $\numDopplerBin$ and $\freqSpacing$ remain constant for a fixed deadline {\eqref{eq:bandwidth_frametime}}.

Fig.~\ref{fig:eval_time_ccdf_plat} shows the CCDF of the processing latency across GPUs based on 10K packets, highlighting the differences in their ability to handle smaller frame sizes.
As the latency deadline increases with larger $\numDopplerBin$, each GPU is evaluated at the \emph{largest} $\numDopplerBin$ value for which its 99.9\% processing latency satisfies the real-time deadline, i.e., the intersection in Fig.~\ref{fig:eval_e2e_time_vs_N_cross_platform}\subref{subfig:eval_e2e_time_vs_N_cross_platform_median}, where $\numDopplerBin = 32$ for RTX 6000 Ada and H200, $\numDopplerBin = 64$ for A100, and $\numDopplerBin = 256$ for Jetson Orin, all with $\numDelayBin=1024$.
Note that all configurations share the same bandwidth and I/Q sampling rate; hence, the selection of $\numDopplerBin$ is based on the channel Doppler spread $\chMaxDopplerFreq$ (and thus the coherence time $\coherenceTime$, Section~\ref{subsec:exp_setup_eval_metrics}) and each GPU's capabilities.

\section{Conclusions}

In this paper, we presented a scalable and practical real-time Zak-OTFS processing system on GPU platforms.
Specifically, we exploited compact matrix operations for Zak transform and channel estimation, and a branchless iterative equalizer.
In addition, the structured sparsity of the channel matrix was derived and leveraged to significantly reduce the complexity of the channel estimation and equalization.
With the insights on hardware-algorithm co-design, our system enabled real-time processing on GPUs that scales up to the grid size of $(16384,32)$ with a deadline of {2.13}\thinspace{\msec}, yielding {491.44}\thinspace{Mbps} and {906.52}\thinspace{Mbps} for QPSK and 16QAM, with $\mathrm{BER} = 0.015\%$ and $7.78\%$ at {25}\thinspace{dB} SNR, respectively.

\bibliographystyle{IEEEtran}
\bibliography{reference}

\end{document}